\documentclass[10pt,letter]{article}

\usepackage{amsmath,amstext,amsgen,latexsym}
\usepackage{amstext,amssymb,amsfonts,latexsym}
\usepackage{theorem}
\usepackage{pifont}

 \newcommand{\bs}{\bigskip}
 \newcommand{\ms}{\medskip}
 \newcommand{\n}{\noindent}
 \newcommand{\s}{\smallskip}
 \newcommand{\hs}[1]{\hspace*{ #1 mm}}
 \newcommand{\vs}[1]{\vspace*{ #1 mm}}

 \newcommand{\real}{\mathbb{R}}
 \newcommand{\nat}{\mathbb{N}}
 \newcommand{\integer}{\mathbb{Z}}

 \newcommand{\field}{\mathbb{F}}

 \newcommand{\ie}{\textrm{i.e.},\hspace*{2mm}}
 \newcommand{\eg}{\textrm{e.g.},\hspace*{2mm}}

 \newcommand{\AAA}{{\cal A}}
 \newcommand{\BB}{{\cal B}}
 \newcommand{\DD}{{\cal D}}

 \newcommand{\KK}{{\cal K}}

\def\bbox{\vrule height6pt width6pt depth1pt}
\theoremstyle{plain}
\theoremheaderfont{\bfseries}
\newtheorem{theorem}{Theorem}[section]
\newtheorem{lemma}[theorem]{Lemma}
\newtheorem{proposition}[theorem]{Proposition}
\newtheorem{corollary}[theorem]{Corollary}

{\theorembodyfont{\rmfamily}
     \newtheorem{definition}[theorem]{Definition}}
{\theorembodyfont{\rmfamily} }
{\theorembodyfont{\rmfamily} }
\newtheorem{claim}{Claim}
\newtheorem{appendixlemma}{Lemma~B.\hs{-1}}
\newenvironment{proof}{\par \noindent
            {\bf Proof. \hs{2}}}{\hfill$\Box$ \vspace*{3mm}}

\newenvironment{proofof}[1]{\vspace*{5mm} \par \noindent
         {\bf Proof of #1.\hs{2}}}{\hfill$\Box$ \vspace*{3mm}}

 \newcommand{\ceilings}[1]{\lceil #1 \rceil}
 \newcommand{\floors}[1]{\lfloor #1 \rfloor}
 \newcommand{\pair}[1]{\langle #1 \rangle}
 \newcommand{\tinycomb}[2]{\left({\tiny \begin{array}{c} #1 \\%
      #2 \end{array} }\right)}
 
 \newcommand{\smatrix}[4]{\left(\begin{array}{cc} #1 &  #2 \\%
      #3  &  #4 \end{array}\right)}
 \newcommand{\comb}[2]{\left( \begin{array}{c} #1 \\%
      #2 \end{array} \right)}    
      
 \newcommand{\bra}[1]{\langle #1 |}
 \newcommand{\ket}[1]{| #1 \rangle}
 \newcommand{\braket}[2]{\langle #1 | #2 \rangle}
 \newcommand{\ketbra}[2]{| #1 \rangle\langle #2 |}

 \newcommand{\qubit}[1]{| #1 \rangle}

 \newcommand{\norm}[1]{\Vert #1 \Vert}

 \newcommand{\prob}{\mathrm{Prob}}
 \newcommand{\legendre}[2]{\left(\frac{#1}{#2}\right)}

\newcommand{\ignore}[1]{}

\newcommand{\EQ}{{\rm EQ}}
\newcommand{\GL}{{\rm GL}}
\newcommand{\SLS}{{\rm SLS}}
\newcommand{\PEQ}{{\rm PEQ}}
\newcommand{\HAD}{{\rm HAD}}

\newcommand{\Pre}{{\rm Pre}}
\newcommand{\ip}[2]{\langle #1,#2 \rangle}

\begin{document}
\pagestyle{plain}
\begin{center}
{\Large {\bf Quantum Hardcore Functions by \s\\
Complexity-Theoretical Quantum List Decoding}}\footnote{An extended abstract appeared in the Proceedings of the 33rd International Colloquium on Automata, Languages and Programming (ICALP 2006), Lecture Notes in Computer Science, Vol.4052 (Part II), pp.216--227. Venice, Italy. July 10--14, 2006.} \bs\bs\\
\begin{tabular}{cc}
{\sc Akinori Kawachi} & {\sc Tomoyuki Yamakami} \ms\\
{\small Graduate School of Information Science and} 
& {\small ERATO-SORST Quantum Computation and Information} \\
{\small  Engineering, Tokyo Institute of Technology} 
& {\small Project, Japan Science and Technology Agency} \\
{\small 2-12-1 Ookayama, Meguro-ku, Tokyo 152-8552, Japan} 
& {\small 5-28-3 Hongo, Bunkyo-ku, Tokyo 113-0033, Japan} \\
\end{tabular}
\bs\\

\end{center}

\bs\bs
\begin{abstract}
Hardcore functions have been used as a technical tool to construct secure cryptographic systems; however, little is known on their quantum counterpart, called {\em quantum hardcore functions}. With a new insight into fundamental properties of quantum hardcores, we present three new quantum hardcore functions for any (strong) quantum one-way function. We also give a {\lq\lq{quantum}\rq\rq} solution to Damg{\aa}rd's question (CRYPTO'88) on a classical hardcore property of his pseudorandom generator, by proving its quantum hardcore property. 
Our major technical tool is the new notion of quantum list-decoding of 
``classical'' error-correcting codes (rather than 
``quantum'' error-correcting codes), which is defined on the platform of computational complexity theory and computational cryptography (rather than information theory). In particular, we give a simple but powerful criterion that makes a polynomial-time computable classical block code (seen as a function) a quantum hardcore for all quantum one-way functions. On their own interest, we construct efficient quantum list-decoding algorithms for classical block codes whose associated quantum states (called codeword states) form a nearly phase-orthogonal basis. 

\ms
\n{\bf Keywords:} quantum hardcore, quantum one-way, quantum list-decoding, codeword state, phase orthogonal, presence, Johnson bound 

\ms
\n{\bf AMS Subject Classifications:} 14G50, 81P68, 94A60

\end{abstract}

\section{From Hardcores to List Decoding}

Modern cryptography heavily relies on computational hardness and pseudorandomness. One of its key notions is a {\em hardcore bit} for a one-way function---a bit that could be determined from all the information available to the mighty adversary but still looks random to any {\lq\lq}feasible{\rq\rq} adversary. A hardcore function transforms the onewayness into pseudorandomness by generating such hardcore bits of a given one-way function. Such a hardcore function is a crucial element of the construction of a pseudorandom generator as well as a bit commitment protocol from a one-way permutation. A typical example is the inner-product-mod-two function  $\GL_x(r)$ of Goldreich and Levin \cite{GL89}, computing  $\pair{x,r}$, the bitwise inner product modulo two, which constitutes a hardcore bit for any (strong) one-way function.\footnote{Literally speaking, this statement is slightly misleading. 
To be more accurate, such a hard-core function concerns only the 
one-way function of the form $f'(x,r)=(f(x),r)$ with $|r|=poly(|x|)$ induced from an arbitrary  
strong one-way function $f$. See, e.g., \cite{Gol01} for a detailed 
discussion.} 
Since $\GL_x(r)$ equals the $r$th bit of 
the codeword $\HAD^{(2)}_x=(\ip{x}{0^n},\ip{x}{0^{n-1}1},\cdots,\ip{x}{1^n})$ of message $x$ of a binary Hadamard code, Goldreich and Levin
essentially gave a polynomial-time list-decoding algorithm for this Hadamard code.
Later, Goldreich, Rubinfeld, and Sudan \cite{GRS95} discussed the 
hardcore property of $q$-ary Hadamard codes.
In the recent literature, list-decoding has kept playing a key role in a general construction of hardcores \cite{AGS03,HMS04}.

Thirteen years later, the ``quantum'' hardcore property (\ie a hardcore property against any feasible ``quantum'' adversary) of $\GL_x(\cdot)$ was shown by Adcock and Cleve \cite{AC02}, who implicitly gave 
a simple and efficient quantum algorithm that recovers $x$ from  the binary 
Hadamard code by exploiting the robust nature of a quantum algorithm of Bernstein and  Vazirani \cite{BV97}. The simplicity of the proof of Adcock and Cleve can be best compared to the original proof of Goldreich and Levin, who employed a rather complicated algorithm with powerful techniques:  self-correction property of the aforementioned Hadamard code and pairwise independent sampling. This highlights a significant role of robust quantum computation in the case of list-decoding (and thus hardcores); however, it has been vastly unexplored until our work except for a quantum decoder of Barg and Zhou \cite{BZ98} for simplex codes. No other quantum hardcore has been proven so far.  The efficiency of robust quantum algorithms with access to biased oracles has been  also discussed in a different context \cite{AIKPY05,BNRW04,HMW03}.

As our main result, we present three new quantum hardcore functions: {\em $q$-ary Hadamard codes} $\HAD^{(q)}$, {\em shifted Legendre symbol codes} $\SLS^{p}$, and {\em pairwise equality codes} $\PEQ$ (see Section \ref{sec:new-hardcore} for their definitions), for any (strongly) quantum one-way function. The first hardcore function is a (nontrivial) extension of the aforementioned result of Adcock and Cleve \cite{AC02}, and the latter two are not yet known to be classical hardcores (see, \eg \cite{GN01}). With regard to $\SLS^{p}$, in particular, we can prove the quantum hardcore property of  Damg{\aa}rd's pseudorandom generator \cite{Dam88}. This gives a ``quantum'' solution to his  question of whether his generator has the classical hardcore property (this is also listed as an open problem in \cite{GN01}). 

Our argument proceeds by relating proving the quantum hardcore property of a given code $C$ (seen as a function) to solving the {\em quantum list-decoding problem} for $C$  via direct access to a {\em quantum-computationally (or quantumly) corrupted codeword}, which is given as a black-box oracle. A quantum list-decoding algorithm  (called a {\em quantum list-decoder}) tries to list all message candidates whose codewords match the quantumly-corrupted codeword within a certain error rate bound. 
Let us first assume that the target function $C$ is not any quantum hardcore for a certain quantum one-way function $f'$, of the form $f'(x,r)=(f(x),r)$, induced from another quantum one-way function $f$. Next, we reduce proving the quantum hardcore property to constructing a polynomial-time quantum list-decoder. Using this list-decoder, we further construct a polynomial-time quantum algorithm that inverts $f'$ with reasonable probability. This  clearly contradicts the one-wayness of $f'$ and hence proves the quantum hardcore property of $C$, as requested. Therefore, our major task of this paper is to present fast quantum list-decoders for the aforementioned three codes. 

Our proof technique exploits the quantum list-decodability of classical error-correcting codes (rather than quantum error-correcting codes). For our purpose, we formulate the notion of {\em complexity-theoretical} quantum list-decoding as a means of message recovery with quantum-computational errors rather than information-theoretical errors, which are usually associated with classical transmission errors. This notion naturally expands the classical framework of list-decoding. Most significantly, our quantum list-decoders tend to be more query-efficient than their classical counterparts; namely, the quantum list-decoders have lower query complexity, which refers to the total number of queries to a given quantumly corrupted codeword. 
For instance, we can build a quantum list-decoder for a $q$-ary Hadamard code $\HAD^{(q)}$ that requires significantly fewer queries than any well-known classical list-decoder for $\HAD^{(q)}$.

Intuitively, a quantumly corrupted codeword expresses the behaviors of (possibly) faulty quantum encoders. In classical list-decoding, a ``classically'' corrupted codeword is generated by a faulty channel as a result of its transmission error. Particularly, it is useful to treat transmission error as a faulty encoding process of messages to codewords when we seek applications of list-decoding in computational complexity. Another key notion of this paper is a useful quantum state, called a {\em ($k$-shuffled) codeword state}, which uses quantum ``phase'' to store the information on a given codeword. Similar states have appeared to play a key role in several quantum algorithms in the recent literature \cite{BV97,DJ92,Gro97,DHI03}.  We then reduce constructing a quantum list-decoder to constructing a {\em quantum codeword-state decoder} (\ie a quantum algorithm that recovers a message $x$ from a codeword state which is given as an input). 
Quantum algorithms of van Dam, Hallgren, and Ip \cite{DHI03} to ``hidden shift'' problems are actually an early example of quantum codeword-state decoders.
In our key lemmas, we show (i) how to generate such a codeword state from {\em any} (even adversarial) quantumly corrupted codeword and (ii) how to convert a codeword-state decoder to a quantum list-decoding algorithm working with a quantumly corrupted codeword. The robust construction made in the course of our proofs also provides a useful  means, known as {\lq\lq{hardness}\rq\rq} reduction, which is often crucial in the security proof of a quantum cryptosystem. Using pretty good measurement \cite{EF01,HW94}, we can present a generic way of proving the quantum list-decodability of a code if the set of its corresponding codeword states forms a {\lq\lq}nearly{\rq\rq} orthogonal basis. This construction method is general and  constructive but not time-efficient. For particular circulant codes, such as $\SLS^{p}$, however, we can give explicitly a quantum list-decoding algorithm. 

Classical list-decodable codes have provided numerous applications in classical computational complexity theory, including proving hardcores for any one-way 
function, hardness amplification, and derandomization (see, \eg \cite{Sud00}). Because our formulation of quantum list-decoding naturally extends classical one, many classical list-decoding algorithms work in our quantum setting as well. This will make our quantum list-decoding a powerful tool in quantum complexity theory and quantum computational cryptography.

\section{New Quantum Hardcore Functions}\label{sec:new-hardcore}

We briefly give the formal definitions to the core concepts of this paper---quantum one-way functions and quantum hardcore functions and then state our main contributions to the emerging field of quantum cryptography. We assume the reader's basic knowledge on quantum computation. Our underlying computation model is quantum Turing machines \cite{BV97,Yam99} and quantum circuits \cite{Yao97}. Informally, we use the term ``quantum algorithm'' to describe a description of a certain unitary operator, possibly together with  a specific projection measurement at the end of a computation. For convenience, the notation $\AAA(x)$ for a quantum algorithm $\AAA$ and an input $x$ denotes a {\em random variable} representing the outcome of the execution of $\AAA$ on input $x$. 

\subsection{Quantum Hardcore Property}

We begin with the notion of {\em quantum one-way functions}, which straightforwardly expands the classical one-way functions introduced first by Diffie and Hellman \cite{DH76} in 1976. 
 Let $\nat$ denote the set of all nonnegative integers.  

\begin{definition}[quantum one-wayness]
Let $\Sigma$ be any alphabet. A function $f$ from $\Sigma^*$ to $\Sigma^{*}$ is called {\em (strongly) quantum one-way} if (i)~there exists a polynomial-time {\em deterministic} algorithm $G$ computing $f$ and (ii)~for any polynomial-time quantum algorithm $\AAA$, 
for any positive polynomial $p$, and for any sufficiently large numbers $n\in\nat$,
\[
 \prob_{x\in\Sigma^n,\AAA}\left[f(\AAA(f(x),1^n)) =f(x)\right] < \frac{1}{p(n)},
\] 
where $x$ is uniformly distributed over $\Sigma^n$ and the subscript $\AAA$ is a 
random variable determined by measuring the final state of $\AAA$ in the 
computational basis. This paper considers only length-regular one-way functions, where a function $f$ mapping $\nat$ to $\nat$ is called {\em length regular} if, for every string $x\in\Sigma^*$, $|f(x)|= \ell(|x|)$ for a certain length function $\ell(n)$.
\end{definition}

Because of the deterministic feature of the above $f$, all quantum one-way functions are indeed classically one-way. For any quantum one-way function $f$, the notation $f'$ denotes the function induced from $f$ by the following randomization scheme: 
$f'(x,r) = (f(x),r)$ for all $x,r\in\Sigma^*$ with $|r|=poly(|x|)$, where the notation $(y,r)$ means the concatenation of $y$ and $r$ following $y$. Notice that 
$f'$ is also a quantum one-way function. Throughout this paper, we deal only with quantum one-way functions of this form, which is  in direct connection to the following notion of quantum hardcore functions.

The notion of a classical hardcore was first discussed by Blum and Micali \cite{BM84} in 1984. A hardcore measures the hardness of predicting the value $h(x)$ from $f(x)$ without knowing $x$ as an explicit input. 
We define a hardcore function $h$ mapping $\Sigma^n$ to $\Sigma^{\ell(n)}$ by the notion of {\em indistinguishability} between $h(x)$ and a truly random variable $z$ over $\Sigma^{\ell(n)}$ rather than the notion of {\em nonapproximability}. 
This is because indistinguishability and nonapproximability define the same hardcore notion as long as the output size of hardcore functions is limited to  $O(\log{n})$ (see, \eg \cite{Gol01}).

\begin{definition}[quantum hardcore]\label{def-strong-hardcore}
Let $\Sigma$ be any alphabet and let $f$ be any length-regular function from $\Sigma^*$ to $\Sigma^*$. A polynomial-time computable function
$h$ with length function $\ell(n)$ is called a {\em quantum hardcore (function)} of $f$ if,
for any polynomial-time quantum algorithm $\AAA$, for any polynomial $p$, and for
any sufficiently large number $n\in\nat$,
\[
 \left|\prob_{x\in \Sigma^n,\AAA}[\AAA(f(x),1^n) = h(x)]-
\frac{1}{|\Sigma|^{\ell(n)}} \right| <  \frac{1}{p(n)},
\]
where $x$ is uniformly distributed over $\Sigma^n$ and the subscript $\AAA$ is a 
random variable determined by measuring the final state of $\AAA$ in the 
computational basis.
\end{definition}

Since every classical randomized algorithm can be translated into a special form of a quantum algorithm, every classical hardcore is also a quantum hardcore. We are mostly interested in the property that a function $h$ becomes a quantum hardcore of {\em any} quantum one-way function $f'$ (of the form $f'(x,r)=(f(x),r)$ for an appropriate quantum one-way function $f$). Succinctly, we refer to this property as the {\em quantum hardcore property} of $h$. 

\ms

As a main theme of this paper, we consider only functions expressed in the forms of block (error-correcting) codes. 
Generally speaking, 
a block (error-correcting) {\em code} is a set of strings of the same length over
a finite alphabet. Each string in a code is indexed by a message and
 is called a {\em codeword}. For our purpose, we are focused on a {\em family of codes}, which is specified by a series $\{(\Sigma_n,I_n,\Gamma_n)\}_{n\in\nat}$ of {\em message space} $\Sigma_n$, {\em index set} $I_n$, and {\em code alphabet} $\Gamma_n$ associated with a length parameter $n$. For convenience, we write $\Sigma^*$ for the set $\bigcup_{n\in\nat}\Sigma_n$.

As standard now in computational complexity theory, we 
view the code $C$ as a function that, for each {\em message length} $n$ (which serves as a {\em basis parameter} in this paper), maps 
$\Sigma_n\times I_n$ to $\Gamma_n$. We sometimes write $C^{(n)}$ to denote the code $C$ restricted to messages of length $n$. Notationally, we set $N(n) = |\Sigma_n|$ and $q(n)=|\Gamma_n|$. It is convenient to assume that $\Sigma_n= (\Gamma_n)^n$ so that $n$ actually represents the {\em length} of a message over $\Gamma_n$. By abbreviating $C(x,r)$ as $C_{x}(r)$, we also treat $C_{x}(\cdot)$ as a function mapping $I_n$ to $\Gamma_n$. Denote by $M(n)$ the {\em block length} $|I_n|$ of each codeword. We simply set $I_n=\{0,1,\ldots,M(n)-1\}$, each element of which can be expressed in $\ceilings{\log_{2}M(n)}$ bits. We freely identify $C_x$ with the vector $(C_{x}(0),C_{x}(1),\cdots,
C_{x}(M(n)-1))$ in the {\em ambient space} $(\Gamma_n)^{M(n)}$ of dimension $M(n)$. We often work on a finite field and it is convenient to regard $\Gamma_n$ as the finite field $\field_{q(n)}$ 
($=\mathrm{GF}(q(n))$) of order $q(n)$, provided that $q(n)$ is a prime power.
The {\em (Hamming) distance} $d(C_x,C_y)$ between two codewords $C_x$ and $C_y$ is the number of non-zero components in the vector $C_x - C_y$. 
In contrast, $\Delta(C_x,C_y)$ denotes the {\em relative (Hamming) distance} $d(C_x,C_y)/M(n)$.
Moreover, the {\em distance} $d(n)$ of a code $C$ of message length $n$ is the minimum distance between any pair of distinct codewords of message length $n$ in $C$. 

The above-described code is simply called a $(M(n),n)_{q(n)}$-code 
(or $(M(n),n,d(n))_{q(n)}$-code if $d(n)$ is emphasized). 
We may drop a length parameter $n$ whenever we discuss 
a set of codewords for a {\lq\lq}fixed{\rq\rq} length $n$; for instance, write $\Gamma$ and $M$ respectively for $\Gamma_n$ and $M(n)$.

\ms

Now, let us present three new quantum hardcore functions, two of which are unknown, at present, to be classical hardcores. These new quantum hardcores are (i) $q$-ary Hadamard codes, (ii) shifted Legendre symbol codes, and (iii) pairwise equality codes. We explain these quantum hardcores as codes and give polynomial-time quantum list-decoding algorithms for them.

\begin{theorem}\label{new-hardcore}
Let $p(n)$ and $q(n)$ be any two functions mapping $\nat$ to the prime numbers with $q(n)\in n^{O(1)}$ and  $n=\ceilings{\log{p(n)}}$. 
The three functions $\HAD^{(q)}$, $\SLS^{p}$, and $\PEQ$, introduced below as classical block codes,  
are all quantum hardcore functions for 
any quantum one-way function $f'$ of the form 
$f'(x,r)=(f(x),r)$ for any $x$ and any $r$ with $|r|=s(|x|)$, where $f$ is an arbitrary quantum one-way function and $s$ is a polynomial.
\begin{enumerate}
\item The {\em $q(n)$-ary Hadamard code} $\HAD^{(q)}$,
whose codeword $\HAD^{(q)}_{x}$ is defined as $\HAD^{(q)}_x(r) = \sum_{i=0}^{2^n-1}x_i\cdot r_i \bmod q(n)$. The distance $d(\HAD^{(q)})$ is $(1-1/q(n))q(n)^n$.
\vs{-1}

\item The {\em pairwise equality code} $\PEQ$ for even numbers $n\in\nat$, which is a $(2^n,n)_2$-code, whose codeword is  
$
 \PEQ_x(r) = \bigoplus_{i=0}^{n/2}\; \EQ(x_{2i}x_{2i+1},r_{2i}r_{2i+1}),
$
where {\rm EQ} denotes the equality predicate (\ie $\EQ(x,y)=1$ if $x=y$ and $0$ otherwise) and $\oplus$ is the bitwise XOR. 
\vs{-1}

\item The {\em shifted Legendre symbol code} $\SLS^{p}$, which is a $(p(n),n)_2$-code with odd prime $p(n)$, whose codeword $\SLS^{p}_x$ is defined by the Legendre symbol\footnote{For any odd prime $p$, let $\left(\frac{x}{p}\right) = 0$ if $p|x$, $\left(\frac{x}{p}\right) = 1$ if $p\!\!\not| x$ and $x$ is a quadratic residue modulo $p$, and $\left(\frac{x}{p}\right) =-1$ otherwise.} as $\SLS^p_x(r)= 1$ if 
$\left(\frac{x+r}{p(n)}\right) =-1$, and $\SLS^{p}_x(r)=0$ otherwise.
\end{enumerate}
\end{theorem}

Earlier, Damg{\aa}rd \cite{Dam88} introduced the so-called {\em Legendre generator}, which takes input $(p(n),x)$ and produces a $q(n)$-bit sequence whose $r$th bit equals $\SLS^{p}_x(r)$  for every index $r\in\field_{p(n)}$, where $p$ is a fixed polynomial. He asked whether his generator possesses the classical hardcore property (which is also listed as an open problem in \cite{GN01}.) Theorem \ref{new-hardcore}(3) proves the {\lq\lq}quantum{\rq\rq} hardcore property of Damg{\aa}rd's generator for any quantum one-way function. 

\section{How can We Prove the Quantum Hardcore Property?}\label{sec:hardcore-property}

We shall outline our argument of proving the quantum hardcore property of a given function. To prove new quantum hardcores, we exploit the notion of quantum list-decoding as a technical tool. Our approach toward list-decoding is, however, {\em complexity-theoretical} in nature rather than information-theoretical. Our main objects of quantum list-decoding are {\lq\lq}classical{\rq\rq} block codes and their codewords, which are  
manipulated in a quantum fashion. 

\subsection{Quantum List Decoding}

In classical list-decoding, we are allowed to access a {\em received word}, which is a {\lq\lq}classically corrupted codeword,{\rq\rq} given as an oracle and our goal is to produce a short list of message candidates that match the received word.
Similar to the classical notion of a received word in coding theory, we introduce our terminology concerning an oracle that represents a {\lq\lq}quantum-computationally{\rq\rq} (or {\lq\lq}quantumly{\rq\rq}) corrupted codeword that produces garbage information of $\ell(n)$ size. 
For an immediate comparison to a quantum case, we use a more conceptual term ``classically corrupted codeword'' instead of the conventional term ``received word'' in the rest of this paper.

Let us formulate a notion of quantum list-decoding and present a key theorem, Theorem \ref{code-hardcore}, which bridges between quantum hardcore functions and quantum list-decoding.

\begin{definition}[quantum-computationally corrupted codeword]\label{corrupted-word}
We say that a unitary operator $\tilde{O}$ 
{\em represents a quantum-computationally (or quantumly) corrupted codeword}  if there exists a function $\ell$ mapping $\nat$ to $\nat$ such that, for any length parameter $n\in\nat$, any index $r\in I_n$, and any element $u\in\Gamma_n$, the operator $\tilde{O}$ satisfies 
\[
\tilde{O}\ket{r}\ket{u}\ket{0^{\ell(n)}} = \sum_{s\in\Gamma_n}\alpha_{r,s}\ket{r}\ket{u\oplus s}\ket{\phi_{r,s}}
\]
for complex numbers $\alpha_{r,s}$ with $\Sigma_{s\in \Gamma_n}|\alpha_{r,s}|^2=1$ for each index $r\in I_n$, and for unit vectors $\ket{\phi_{r,s}}$ in a $2^{\ell(n)}$-dimensional Hilbert space, depending only on $(r,s)$, where $\oplus$ is the bitwise XOR.
Here, the parameter $\ell(n)$ indicates the size of {\em garbage information} produced by $\tilde{O}$. Notice that $\tilde{O}$ is a unitary operator acting on a Hilbert space spanned by the elements of $\bigcup_{n\in\nat}(I_n\times\Gamma_n\times\{0,1\}^{\ell(n)})$. For convenience, we identify a quantumly corrupted codeword with its representing oracle and we simply call $\tilde{O}$ a quantumly corrupted codeword.
\end{definition}

With the above notion, we shall describe a quantum version of a {\em classical list-decoding problem}. Let us recall that our target function $C(x,r)$ is always expressed as an $(M(n),n,d(n))_{q(n)}$-code family $C=\{C_x\}_{x\in\Sigma^*}$.

For the formulation of the quantum list-decoding problem, we introduce a notion of {\em presence}, which is a quantum analogue of a ``closeness''-scale between a codeword and its quantumly corrupted codeword. More precisely, fix $n\in\nat$ and $x\in\Sigma_n$ and consider the entity $(1/M(n))\sum_{r\in I_n}|\alpha_{r,C_x(r)}|^2$ (using the terms given in Definition \ref{corrupted-word}). In comparison, let us review classical list decoding. For any given oracle $O$ that represents a {\em classically corrupted codeword} and for any error bound $\varepsilon$, a classical list decoder tries to output a short list consisting of all messages $x$ for which the probability, over $r\in I_n$, that $O(r)$ equals $C_x(r)$ is at least  $1-\varepsilon$ 
(namely, $\prob_{r\in I_n}[O(r)=C_x(r)]\geq 1-\varepsilon$). 
By setting $p_{r,s} = 1$ if $\tilde{O}(r)=s$ and $0$ otherwise,
the behavior of $\tilde{O}$ can be viewed in a style of unitary operation  as $\tilde{O}\ket{r}\ket{0} =\sum_{r\in I_n}p_{r,s}\ket{r}\ket{s}$. The aforementioned entity
$(1/M(n))\sum_{r\in I_n}|\alpha_{r,C_x(r)}|^2$ equals the probability  $\prob_{r\in I_n}[\tilde{O}(r)=C_x(r)]$ in a classical setting. 
For our convenience, we name this entity 
the {\em presence} of $C_x$ in $\tilde{O}$ and use the notation  $\Pre_{\tilde{O}}(C_x)$ to describe it. 

Our task of quantum list-decoding of the code $C$ is, given a quantumly corrupted codeword $\tilde{O}$ for $C$ and a size parameter $n$ (as well as garbage size $\ell(n)$ of $\tilde{O}$), to produce with success probability at least $\delta(n)$ a list of candidates $x$, all of which satisfy the condition $\Pre_{\tilde{O}}(C_x)\geq 1/q(n)+\varepsilon(n)$, where $\varepsilon(n)$ is an error bias parameter and $\delta(n)$ is a confidence parameter. 
For our application of quantum list-decoding to quantum hardcore, 
we need to deal only with quantumly corrupted codewords $\tilde{O}$ of polynomial garbage size $\ell$. 
A ``quantum list decoding algorithm'' (or a ``quantum list decoder'') means a procedure of carrying out this task with the two parameters $\varepsilon$ 
and $\delta$. 

Now, we formally define this notion as follows.

\begin{definition}[quantum list decoder]
Let $C$ be any $(M(n),n)_{q(n)}$-code family, let $\varepsilon(n)$ be any error bias parameter and let $\delta(n)$ be any confidence parameter. Moreover, let $\ell$ be any polynomially-bounded function from $\nat$ to $\nat$. 
A {\em quantum list-decoding algorithm} for $C$ with respect to $(\ell,\varepsilon,\delta)$ (or an {\em $(\ell,\varepsilon,\delta)$-quantum list-decoder}) is a  quantum algorithm (\ie a unitary operator) $\DD$ that solves with success probability at least $\delta(n)$ the following quantum list-decoding problem:

\begin{itemize}\vs{-1}
\item[] {\sc Input:} a message length $n$.
\vs{-2}
\item[] {\sc Implicit Input:} an oracle $\tilde{O}$ representing a quantumly corrupted codeword of garbage size $\ell(n)$.
\vs{-2}
\item[] {\sc Output:} a list of messages including all messages $x\in \Sigma_n$ such that 
$
 \Pre_{\tilde{O}}(C_x) \geq 1/q(n) + \varepsilon(n);
$ 
in other words, codewords $C_x$ have {\lq\lq}slightly{\rq\rq} higher presence in $\tilde{O}$ than the average. For convenience, we call such a list a {\em valid list}.
\end{itemize}
\s

If $\DD$ further runs in time polynomial in $(n,1/\varepsilon(n),\log_{2}[1/(1-\delta(n))])$ (if $\delta(n)=1$ then  we treat $1/(1-\delta(n))$ as $1$ for notational convenience), 
it is called a {\em polynomial-time quantum 
list-decoding algorithm} for $C$ with respect to $(\ell,\varepsilon,\delta)$. We also say that $C$ is {\em $(\ell,\varepsilon,\delta)$-quantum list-decodable} if $C$ has an $(\ell,\varepsilon,\delta)$-quantum list-decoder.
\end{definition}

Now, we are ready to give our key theorem, Theorem \ref{code-hardcore}, which serves as a driving force to develop a theory of quantum list decoding in the subsequent sections. Notably, Sections \ref{sec:list-decoding}--\ref{sec:circulant} will be devoted to the construction of quantum list-decoders for each given quantum hardcore candidate. The first step is to establish a generic technique of constructing quantum list-decoders for ``well-behaved'' classical block codes.

\begin{theorem}\label{code-hardcore}
Let $C=\{C_x\}_{x\in\Sigma^*}$ be any $(M(n),n,d(n))_{q(n)}$-code with a message space $\Sigma^*=\bigcup_{n\in\nat}\Sigma_n$, 
which
is polynomial-time computable, 
where $\log_{2}{M(n)}\in n^{O(1)}$ and $\log_{2}q(n)\in 
n^{O(1)}$. 
If, for any noticeable\footnote{A function $\mu$ from $\nat$ to $\real$ is said to be {\em noticeable} if there exists a positive polynomial $p$ such that $\mu(n)\geq 1/p(n)$ for any sufficiently large number $n\in\nat$.}
function $\varepsilon(n)$ and any polynomially-bounded function $\ell(n)$, there exist a noticeable function $\delta(n)$ and a 
polynomial-time $(\ell,\varepsilon,\delta)$-quantum list-decoder for 
 $C$, then 
$C(x,r)$ is a quantum hardcore for any quantum one-way 
function $f'$ of the form $f'(x,r)=(f(x),r)$ with 
$|x|=\ceilings{\log_{2}{N(n)}}$ and 
$|r|=\ceilings{\log_{2}{M(n)}}$, where $N(n)=|\Sigma_n|$, and therefore $C$ satisfies the quantum hardcore property. 
\end{theorem}

\subsection{Quantumly Corrupted Codewords}

Through this subsection to the next two subsections, we shall explain how to prove Theorem \ref{code-hardcore} with a succinct justification of our notions of quantumly corrupted codeword, presence, and quantum list-decoding. 
 
Now, let $C=\{(\Sigma_n,I_n,\Gamma_n)\}_{n\in\nat}$ be any $(M(n),n,d(n))_{q(n)}$-code family, where $C$ is viewed as a collection of functions mapping $\Sigma_n\times I_n$ to $\Gamma_n$ for each length parameter $n\in\nat$. For simplicity, assume that $p(n)=\ceilings{\log_{2}M(n)}$ for a certain polynomial $p$. For every quantum one-way function $f$, we wish to prove that this function $C(x,r)$ is indeed a quantum hardcore for the induced quantum one-way function $f'$ of the form $f'(x,r)=(f(x),r)$ with $|r|=p(|x|)$. 
For simplicity, we assume that all the elements in $I_n$, $\Sigma_n$, and $\Gamma_n$ are expressed in binary using an appropriate, simple, and easy encoding scheme. In the following argument, we fix an arbitrary quantum one-eay function. To lead to a desired contradiction, we first assume to the contrary that there exists a polynomial-time quantum algorithm $\AAA$ that approximates $C_{x}(r)$ from input $(f'(x,r),1^n)$ with success probability, over $(x,r)\in\Sigma_n\times I_n$, at least $1/q(n)+\varepsilon(n)$ (where $\varepsilon(n)$ is a certain {\em noticeable function}); that is, $\prob_{\AAA,(x,r)}[\AAA(f'(x,r),1^n) = C_x(r)]\geq 1/q(n)+\varepsilon(n)$. Thus, there are at least an $\varepsilon(n)/2$-fraction of $|\Sigma_n|$ elements $x\in\Sigma_n$ satisfying that $\AAA(f'(x,r),1^n)$ outputs $C_x(r)$ with probability at least $1/q(n)+\varepsilon(n)/2$. Meanwhile, we fix such an $x$ and let $y=f(x)$. 
The final configuration of the quantum algorithm 
$\AAA$ on input $((y,r),1^n)$, where $r\in I_n$, 
can be assumed to be of the form: 
\[
\alpha_{y,r,C_x(r)}\ket{r}\ket{C_x(r)}\ket{\phi_{y,r,C_x(r)}}
 + \sum_{s\in\Gamma_n-\{C_x(r)\}}\alpha_{y,r,s}\ket{r}\ket{s}\ket{\phi_{y,r,s}}
\]
for certain amplitudes $\alpha_{y,r,s}$ and ancilla quantum states $\ket{\phi_{x,r,s}}$ of $\ell(n)$ qubits, where the second register corresponds to the output of the algorithm, where $\ell(n)$ is a polynomially-bounded function. 

For the fixed string $y$, we pay our attention to the (restricted) algorithm $\AAA_{y}(\cdot) =_{def}\AAA(y,\cdot)$.
An implicit input to our quantum list-decoder is a quantumly corrupted codeword $\tilde{O}_{\AAA_y}$, of garbage size $\ell(n)$, defined by a certain unitary transformation that satisfies the following necessary condition:
\[
\tilde{O}_{\AAA_y}\ket{r}\ket{u}\ket{0^{\ell(n)}} = \sum_{s\in\Gamma_n}\alpha_{y,r,s}\ket{r}\ket{u\oplus s}\ket{\phi_{y,r,s}} 
\]
for every pair $(r,u)$ of strings. This oracle $\tilde{O}_{\AAA_y}$ describes   
{\em computational error} (not transmission error) occurring during the computation of $C_x$ by the (possibly) faulty quantum algorithm $\AAA$. 
This type of erroneous quantum computation is similar to the computational 
errors (e.g., \cite{AC02,AIKPY05,AS04,HMW03}) dealt with in quantum computational 
cryptography and quantum algorithm designing. Notice that the amplitudes $\{\alpha_{y,r,s}\}_{r,s,}$ in $\tilde{O}_{\AAA_y}$ satisfy that $\sum_{s\in\Gamma_{n}}|\alpha_{y,r,s}|^2=1$ for each index $r\in I_n$. 
Since $\tilde{O}_{\AAA_y}$ is a unitary operation, its inverse $\tilde{O}_{\AAA_y}^{-1}$ can be uniquely defined. 

We can freely access $\tilde{O}_{\AAA_y}$ (as well as $\tilde{O}^{-1}_{\AAA_y}$) by simply invoking a query, using three registers containing $(r,u)$. Upon an oracle call, the oracle is automatically applied to the three registers and all the contents of these registers are modified at the cost of unit time.

\subsection{Presence and List Size}

To lead to our desired contradiction, we need to invert the function $f(x)$ by extracting $x$ from the aforementioned quantumly corrupted codeword $\tilde{O}_{\AAA_y}$ in time polynomial in $|x|$. 

Before proceeding our proof further, we make a close look at the presence notion. For our quantumly corrupted codeword $\tilde{O}_{\AAA_y}$ for a target codeword $C_x$, its presence $\Pre_{\tilde{O}_{\AAA_y}}(C_x)$, which is $(1/M(n))\sum_{r\in I_n}|\alpha_{y,r,C_x(r)}|^2$, coincides with the probability that the algorithm $\AAA$ successfully computes $C_x(r)$ from $(y,r)$.  {}From our assumption, it holds that $\Pre_{\tilde{O}_{\AAA_y}}(C_x)\geq 1/q(n)+\varepsilon(n)$. 

{}From a slightly different view point, we can argue that the presence notion is indeed an extension of relative (Hamming) distance. This will be used in the proof of Lemma \ref{johnson-bound}. Let $v$ denote a classically corrupted codeword. We can view $v$ as a binary vector in the $q(n)M(n)$-dimensional space, in which the $r$th block $v[r]$ of $v$ is of the form $0^{i-1}10^{q(n)-i}$ for a certain index $i\in[q(n)]$, where $r\in{M(n)}$. Using this new representation, the relative (Hamming) distance between two classically corrupted codewords $v$ and $w$ equals the $\ell_1$-norm $\|v-w\|_1=\sum_{r\in[M(n)]}\|v[r]-w[r]\|$. Similarly, for a quantumly corrupted codeword $v_{\tilde{O}_{\AAA_y}}$ that the oracle $\tilde{O}_{\AAA_y}$ represents, $v_{\tilde{O}_{\AAA_y}}$ can be viewed as the real vector in the $q(n)M(n)$-dimensional space, in which $v_{\tilde{O}_{\AAA_y}}[r]$ is  $(|\alpha_{r,0}|^2,|\alpha_{r,1}|^2,\ldots,|\alpha_{r,q(n)-1}|^2)$. The presence $\Pre_{\tilde{O}_{\AAA_y}}(C_x)$ now indicates the $\ell_1$-norm between $v_{\tilde{O}_{\AAA_y}}$ and a codeword $C_x$, extending the classical notion of distance. We then obtain $\|v_{\tilde{O}_{\AAA_y}} - C_x\|_{1}\geq 1/q(n)+\varepsilon(n)$. Therefore, if we can find all elements $x'$ satisfying $\Pre_{\tilde{O}_{\AAA_y}}(C_{x'})\geq 1/q(n)+\varepsilon(n)$, at least one of them must satisfy $f(x')=y$. 

Because no quantum list-decoder can output a valid list of super-polynomial size in polynomial time for the quantumly corrupted codeword $\tilde{O}_{\AAA_y}$, there is an important question to answer: how many messages $x$ satisfy the required inequality $\Pre_{\tilde{O}_{\AAA_y}}(C_x) \geq 1/q(n) + \varepsilon(n)$? 
We want to show an upper bound of the number of codewords that 
have relatively high presence in 
a given quantumly corrupted word.
For our proof, we employ a proof method of Guruswami and Sudan \cite{GS00b}, who gave a $q$-ary extension of {\em Johnson bound} using a geometric
method. 

\begin{lemma}\label{johnson-bound}
Let $n$ be any message length. Let $\varepsilon(n)$, $q(n)$, $d(n)$, and $M(n)$ satisfy that $\varepsilon(n) > k(n)=_{def}\left(1 - 1/q(n) \right) \sqrt{1 -  d(n)/M(n) \left( 1 + 1/(q(n)-1)\right) }$. For any $(M(n), n, d(n))_{q(n)}$-code $C$ and
for any given quantumly corrupted codeword $\tilde{O}$, there are at most 
{\small
\[
J_{\varepsilon,q,d,M}(n)=_{def} \min\left\{ M(n)(q(n) - 1), 
\frac{d(n) \left(1 - 1/q(n)\right)}{M(n)\varepsilon^2(n) + \left(1 - 1/q(n) \right)\left[ d(n) - M(n)\left(1 - 1/q(n)\right) \right]} \right\}
\] 
}
\!\!messages $x\in\Sigma_n$ such that $\Pre_{\tilde{O}}(C_x) \geq 1/q(n)+ \varepsilon(n)$. If $\varepsilon(n) = k(n)$, then
the above bound can be replaced by $2M(n)(q(n) - 1) - 1$.
\end{lemma}

The proof of Lemma \ref{johnson-bound} is in essence a simple modification of the proof in \cite{GS00b}; however, for completeness, we include the proof of the lemma in Appendix. As a quick example, we present the value $J_{\varepsilon,q,d,M}(n)$ for a  $(q^n,n,q^n-q^{n-1})_q$ Hadamard code. 
 
\paragraph{Example: Hadamard Codes.}
Consider an $(M(n),n,d(n))_{q(n)}$ Hadamard code $\HAD^{(q)} = \{\HAD^{(q)}_x\}_{x\in\Sigma^*}$ with $M(n) = q(n)^n$ and $d(n) = \left(1 - 1/q(n)\right) M(n)$. Assume that our bias parameter $\varepsilon$ is non zero (\ie $\varepsilon(n) > 0$ for all $n\in\nat$). Lemma \ref{johnson-bound} guarantees that, for any quantumly corrupted codeword $\tilde{O}$, the number of codeword candidates that satisfy the inequality $\Pre_{\tilde{O}}(\HAD^{(q)}_x)\geq 1/q(n)+\varepsilon(n)$ is
at most
\[
\frac{d(n) \left( 1 - \frac{1}{q(n)} \right)}{M(n) \varepsilon(n)^2 
+ \left(1 - \frac{1}{q(n)} \right) \left[ d(n) - M(n)\left( 1 - \frac{1}{q(n)} \right) \right]}
 = \left( 1 - \frac{1}{q(n)} \right)^2 \cdot \frac{1}{\varepsilon(n)^2}.
\]
In particular, if there exists a positive polynomial $p$ satisfying $\varepsilon(n) \geq 1/p(n)$ for all $n\in\nat$, there are only at most
 $\left( 1 - 1/q(n)\right)^2 p(n)^2$ codeword candidates. 
 
\subsection{Applying a Quantum List Decoder}\label{sec:application-oneway}
 
Let us return to our proof of Theorem \ref{code-hardcore}. By the premise of the theorem, there exist a noticeable function $\delta$ and a polynomial-time $(\ell,\varepsilon,\delta)$-quantum list-decoding algorithm $D$ for $C_x(\cdot)$ with certain noticeable probability, say $\delta(n)$. 

For simplicity, we assume that, for each oracle access, our quantum list-decoder $D$ uses its last three registers $\qubit{r}\qubit{u}\qubit{t}$, in which the last register holds a quantum state of polynomially-many qubits. 

Since the garbage size $\ell(n)$ of $\tilde{O}_{\AAA_y}$ is polynomially bounded, we can assume that, after an oracle call, the oracle $\tilde{O}_{\AAA_y}$ (or its inverse $\tilde{O}_{\AAA_y}^{-1}$) is automatically applied to the last three registers of $D$ with a unit cost of time. For convenience, the last register is assumed to hold only $0$s at the beginning of the computation.   

With oracle access to this oracle $\tilde{O}_{\AAA_y}$ (as well as its inverse $\tilde{O}^{-1}_{\AAA_y}$), our quantum list-decoder $D$ can produce with probability at least $\delta(n)$ all possible candidates $x'$ that have the required presence $\Pre_{\tilde{O}_{\AAA_y}}(C_{x'})$ of at least $1/q(n)+\varepsilon(n)/2$. As seen before, at least one of these candidates lies in the pre-image $f^{-1}(y)$. Since we can check whether $f(x')=y$ in polynomial time, it suffices for us to output one of such elements $x'$. This implies that, for at least $(\varepsilon(n)/2)|\Sigma_n|$ elements $x$, we can find $x'$ in polynomial time such that $f(x')=f(x)$. Our quantum list-decoder $D$ therefore gives rise to a polynomial-time quantum algorithm that inverts $f$ on a noticeable fraction of inputs $x$ with noticeable probability. This clearly contradicts the quantum one-wayness of $f$. 

Therefore, the quantum hardcore property holds for $C$ and this completes the proof of Theorem \ref{code-hardcore}. 

\section{Key Roles of Quantum Codeword-State Decoders}\label{sec:list-decoding}

How can we prove the quantum hardcore property of our target quantum hardcore candidates? As outlined in Section \ref{sec:hardcore-property}, our goal is to construct a polynomial-time quantum list-decoder for each of the target candidates. 
Theorem \ref{code-hardcore} gives a sufficient condition to prove the quantum hardcore property of any given code $C$. It is therefore enough for us to design a quantum algorithm that solves the quantum list-decoding problem for $C$ with high probability in polynomial time. Our task in this paper is now fixed to find a generic way to construct a polynomial-time quantum list-decoder for a wide range of classical block codes. Notice that it seems hard to design classically such list-decoding algorithms. 
Let us introduce a central notion of {\em $k$-shuffled codeword states}, which is a unique quantum state encoding all information on a given codeword.

\subsection{Quantum Codeword States}

Hereafter, we assume the basic arithmetic operations (multiplication, addition, subtraction, division, etc.) on a finite field $\field_q$ of order $q$. When $q$ is a prime number, $\field_q$ can be identified with  the integer ring $\integer/q\integer$ whose elements are written as $0,1,2,\ldots,q-1$. For convenience, let $\field_q^{+}$ stand for $\field_q-\{0\}$. The notation $\integer_{q}$ denotes the (finite) additive group whose elements are $0,1,2,\ldots,q-1$. Moreover, we write $[m,n]_{\integer}=\{m,m+1,m+2,\ldots,n\}$ for any two integers $m,n\in\nat$ with $m\leq n$ and, in particular, let $[q] = [1,q]_{\integer}$ for any integer $q\geq1$ in the rest of this paper. Finally, we denote by $\omega_q$ the complex number $e^{2\pi \iota/q}$, where $e$ is the base of natural logarithms and $\iota=\sqrt{-1}$.

\begin{definition}[shuffled codeword state]\label{codeword-state}
\sloppy Let $C=\{C_x\}_{x\in\Sigma^*}$ be any $(M(n),n)_{q(n)}$-code family with a message space $\Sigma^*=\bigcup_{n\in\nat}\Sigma_n$ and a series $\{I_n\}_{n\in\nat}$ of index sets. Let $k$ be any element in $\field^{+}_{q(n)}$. A {\em $k$-shuffled codeword state} for the codeword $C_{x}$ that encodes a message ${x}\in \Sigma_n$ is the quantum state 
\[
\qubit{C_{x}^{(k)}} = \frac{1}{\sqrt{M(n)}}\sum_{{r}\in I_{n}}\omega_{q(n)}^{k\cdot C_{{x}}({r})}\qubit{{r}}. 
\] 
In particular, when $k=1$, we use the simplified notation $\qubit{C_{{x}}}$ for $\qubit{C_{{x}}^{(1)}}$.
\end{definition}

A robust nature of quantum computation enables us to prove that, as long as  we have a decoding algorithm $\AAA$ from a shuffled codeword state, we can construct a quantum list-decoder by calling $\AAA$ 
as a black-box oracle. The notion of such codeword states plays a central role as our technical tool in proving new quantum hardcores. 

The reader may be aware that our notion of codeword states is not anew; the codeword states for certain binary codes have already appeared implicitly in several important  
quantum algorithms. For instance, Grover's search algorithm \cite{Gro97} produces such a codeword state after the first oracle call. In the quantum algorithms of Bernstein and Vazirani \cite{BV97}, of Deutch and Jozsa \cite{DJ92}, and of van Dam, Hallgren, and Ip \cite{DHI03},   
such codeword states are generated to obtain their results. All these quantum algorithms hinge at generating codeword states.

Our task is to recover $x$ with reasonable probability from each $k$-shuffled codeword state $\qubit{C_x^{(k)}}$. Any quantum algorithm that completes this task is succinctly called a {\em codeword-state decoder}. We formally define the quantum codeword-state decoders.

\begin{definition}[quantum codeword-state decodability]
Let $\eta\in[0,1]$. A classical $(M(n),n)_{q(n)}$-code family is said to be {\em $\eta$-quantum codeword-state decodable} if there exists a quantum algorithm that, on input $n\in\nat$ and $k\in\field_{q(n)}^{+}$ as well as  $\qubit{C^{(k)}_{x}}$, recovers $x$ with success probability at least $\eta$. Such an algorithm is simply called an {\em $\eta$-quantum codeword-state decoder}.
\end{definition}

Let us prove the following key theorem, which helps us convert quantum codeword states into quantum list-decoders. The following theorem, Theorem  \ref{list-decoding}, which is general but slightly technical, shows how we can convert a codeword-state decoder $\DD$ for a given code $C$ into a quantum list-decoder $\BB$ for $C$ that produces a valid list of appropriate size. Under a certain condition, we can make this quantum list-decoder to run in polynomial time. 

Recall the definition of $J_{\varepsilon,q,d,M}(n)$ given in Lemma \ref{johnson-bound}. Note that 
$J_{\varepsilon,q,d,M}(n)\leq 2M(n)q(n)$. 
In the following, ``$e$'' stands for the base of the natural logarithm. 

\begin{theorem}\label{list-decoding}
Let $C=\{C_x\}_{x\in\Sigma^*}$ be any $(M(n),n,d(n))_{q(n)}$-code.  
Let $\varepsilon$ and $\delta$ be any two nonnegative functions with $0\leq \varepsilon(n)\leq 1$ and $0\leq \delta(n)< 1$ for all numbers $n\in\nat$. 
Let $\ell$ be any polynomially-bounded function from $\nat$ to $\nat$. If there exists a $(1-\nu(n))$-quantum codeword-state decoder $\DD$ for $C$  with $1-\nu(n)>\sqrt{1-\eta_{\varepsilon}(n)^2}$ for a certain  function $\nu(n)$ from $\nat$ to $[0,1]$, then there exists an  $(\ell,\varepsilon,\delta)$-quantum list-decoder $\BB$ for $C$ with oracle access to $\tilde{O}$ such that $\BB$ produces a list of size at most 
\[
\left\lceil \frac{q(n)-1}{1-\nu(n)-\sqrt{1-\eta_{\varepsilon}(n)^2}} 
\left(\log_{e}{J_{\varepsilon,q,d,M}(n)}+\log_{e}\frac{1}{1-\delta(n)}\right) \right\rceil,  
\]
\sloppy where $\eta_{\varepsilon}(n) = (q(n)/(q(n)-1))\varepsilon(n)$ and the query complexity (\ie the total number of queries to $\tilde{O}$ 
as well as $\tilde{O}^{-1}$) is at most twice as many as the list size. Moreover, letting $\sigma(n)=1-\nu(n)-\sqrt{1-\eta_{\varepsilon}(n)^2}$, if $\DD$ runs in time polynomial in $(n,q(n),\log_{2}M(n))$ and $\sigma$ is a positive-valued function whose reciprocal (\ie $1/\sigma(n)$) is polynomially-bounded in $(n,q(n),\log_{2}M(n),1/\varepsilon(n),\log_{2}(1/(1-\delta(n))))$ from above, 
 then $\BB$ runs in time polynomial in $(n,q(n),\log_{2}M(n),1/\varepsilon(n),\log_{2}(1/(1-\delta(n))))$. 
\end{theorem}

Combining Theorem \ref{list-decoding} together with Theorem \ref{code-hardcore}, we can establish a direct connection between the existence of polynomial-time quantum codeword-state decoder for a code $C$ and the quantum hardcore property of $C$.

\begin{corollary}\label{negligible-code}
Let $s$ be any negligible function mapping 
$\nat$ to the interval $[0,1]$. 
Let $C$ be any $(M(n),n)_{q(n)}$-code family with $\log_{2}M(n)\in n^{O(1)}$ and $q(n)\in n^{O(1)}$. Let $\delta(n)=1-2^{-n}$ for every number $n\in\nat$. If there exists a polynomial-time $(1-s)$-quantum codeword-state decoder for $C$, then, for any noticeable function $\varepsilon$ and any polynomially-bounded function $\ell$, there exists a polynomial-time $(\ell,\varepsilon,\delta)$-quantum list-decoder for $C$. Hence, $C$ satisfies the quantum hardcore property.
\end{corollary}

\begin{proof}
Let $s$ be any negligible function and let $\DD$ be a polynomial-time $(1-s)$-quantum codeword-state decoder for an $(M(n),n)_{q(n)}$-code family $C$. Define $\delta(n)= 1 - 2^{-n}$ for any $n\in\nat^{+}$. Note that $\log_{2}(1/(1-\delta(n))) = \log_{2}2^n = n$. 
Let $\varepsilon$ be any noticeable function. Let $\eta_{\varepsilon}(n) = (q(n)/(q(n)-1))\varepsilon(n)$. By the definition of noticeability, there is an appropriate positive polynomial $p'$ such that $\varepsilon(n)\geq 1/p'(n)$ for any sufficiently large $n\in\nat$.  
To apply Theorem \ref{list-decoding}, we need to show that $\sigma(n) = (1-s(n))-\sqrt{1-\eta_{\varepsilon}(n)^2}$ is a noticeable function (in $n$), because the functions $\log_{2}M(n)$, $q(n)$, $1/\varepsilon(n)$, and $\log_{2}(1/(1-\delta(n)))$ are all polynomially bounded in $n$. Fix any sufficiently large number $n$ in $\nat$ so that the following argument always holds. Since $s$ is a negligible function, it follows that $s(n)\leq 1/4p'(n)^2$. Using the inequality $\sqrt{1-x}\leq 1- x/2$, we obtain 
\[
\sigma(n) \geq 1 - s(n) - \sqrt{1 - \varepsilon(n)^2}
\geq 1 - \frac{1}{4p'(n)^2} - \left(1 - \frac{1}{2p'(n)^2}\right) =  \frac{1}{4p'(n)^2}.
\]
Since $n$ is arbitrary, $\sigma$ is clearly a noticeable function, as requested. By Theorem \ref{list-decoding}, we then obtain an $(\ell,\varepsilon,\delta)$-quantum list-decoder for $C$ running in time polynomial in $n$.
\end{proof}

\subsection{Proof of Theorem \ref{list-decoding}}

We shall prove our key theorem, Theorem \ref{list-decoding}. 
The lemma will be proven in two stages. In the first stage, we present how to generate a quantum state $\qubit{C_x}$ from each quantumly corrupted codeword $\tilde{O}$ for $C_x$. In the second stage, we show how to list-decode $x$ from $\qubit{C_x}$, using a given codeword-state decoder for $C$. These stages give a desired quantum list-decoder for $C$.

As the first step, we shall show how to generate the $k$-shuffled codeword state $\ket{C_x^{(k)}}$ for each $q$-ary codeword $C_x$ with oracle accesses to a quantumly corrupted codeword $\tilde{O}$. 
It is rather straightforward to generate the quantum state $\qubit{C_x}$ from the oracle $O_{C_x}$
that represents $C_x$ {\em without} any corruption (behaving as the {\lq\lq{standard}\rq\rq} oracle). Using $\tilde{O}$, however, it seems difficult to produce $\qubit{C_x}$ with relatively high probability. In Lemma \ref{kappa}, loosely speaking, for a certain constant $k>0$, we can produce in polynomial time a quantum state of the form $\qubit{k}\qubit{C^{(k)}_{x}}\qubit{\tau}$ from an initial state of the form $\qubit{k}\qubit{0^m}\qubit{0}\qubit{0^{\ell}}$ with relatively good probability (if the presence $\Pre_{\tilde{O}}(C_x)$ is far away from $1/q(n)$). Moreover, we claim that there exists a ``generic'' quantum algorithm that generates $\qubit{k}\qubit{C^{(k)}_{x}}\qubit{\tau}$ zfor any $q$-ary code $C$. 

\begin{lemma}\label{kappa}
Let $C$ be any $(M(n),n)_{q(n)}$-code family with a message space $\Sigma^*=\bigcup_{n\in\nat}\Sigma_n$, where $q(n)$ is a prime number for every $n\in\nat$. Let $m$ be any function from $\nat$ to $\nat$. There exists a quantum algorithm $\AAA$ that, for any message length $n\in\nat$, for any quantumly corrupted codeword $\tilde{O}$ with garbage size $\ell(n)$, for any message ${x}\in\Sigma_n$, and for any $k\in \field^{+}_{q(n)}$, generates the quantum state 
\[
\qubit{\psi_k} = \kappa_{{x}}^{(k)}\qubit{k}\qubit{C_{{x}}^{(k)}}\qubit{\tau} + \qubit{\Lambda_{{x}}^{(k)}}
\] 
from the initial state $\qubit{\psi_k^{(0)}} = \qubit{k}\qubit{0^{\ceilings{\log_{2}{M(n)}}}} \qubit{0}\qubit{0^{\ell(n)}}$ with only two queries to $\tilde{O}$ and $\tilde{O}^{-1}$, where $\qubit{\tau}$ is a fixed basis vector, and $\kappa_{{x}}^{(k)}$ is a complex number, and $\qubit{\Lambda_{{x}}^{(k)}}$ is a vector satisfying $(\bra{k}\bra{C_{{x}}^{(k)}}\bra{\tau})\qubit{\Lambda_{{x}}^{(k)}}=0$ with the following condition: for every $x\in\Sigma_n$, there exists an element $k\in \field^{+}_{q(n)}$ with  the inequality $|\kappa_{{x}}^{(k)}| \geq  (q(n)/(q(n)-1))\left| \Pre_{\tilde{O}}(C_x) - 1/q(n)  \right|$. Moreover, $\AAA$ runs in time polynomial in $(n,\log_{2}q(n),\log_{2}M(n))$.
\end{lemma}

Lemma \ref{kappa} provides a generic way of generating a $k$-shuffled codeword state $\qubit{C^{(k)}_{x}}$ from $\tilde{O}$. When $q=2$, the bound of $|\kappa_{x}^{(1)}|$ in Lemma \ref{kappa} matches the bound of Adcock and Cleve \cite{AC02}. 

Now, we give the proof of Lemma \ref{kappa}. Notice that the lemma is true for any $q$-ary code. The binary case ($q=2$) was discussed implicitly in \cite{AC02}; however, our argument for the general $q$-ary case is more involved because of our {\lq\lq}$k$-shuffledness{\rq\rq} condition. 

\begin{proofof}{Lemma \ref{kappa}}
Since $q(n)$ is a prime number, we use $\{0,1,2,\ldots,q(n)-1\}$ as the elements of $\field_{q(n)}$. We assume the premise of the theorem. Let $C$ be any $(M(n),n)_{q(n)}$-code family with message space $\Sigma^*=\bigcup_{n\in\nat}\Sigma_n$, index sets $\{I_n\}_{n\in\nat}$, and code alphabets $\{\Gamma_n\}_{n\in\nat}$. Note that $M(n)=|I_n|$. Let $\tilde{O}$ be any quantumly corrupted codeword of garbage size $\ell(n)$ for $C$, where $\ell$ is an arbitrary polynomially-bounded function. First, we describe our quantum codeword-state generation algorithm $\AAA$ in detail. Fix $n\in\nat$,  ${x}\in\Sigma_n$, and $k\in\field^{+}_{q(n)}$ in the following description. For simplicity, we drop the script ``$n$'' and also let $m= \ceilings{\log_{2}{M}}$.

\vspace{5mm}
\hrule\vspace{3mm}
\n{\sc Quantum Algorithm $\AAA$:}
\begin{enumerate}\ms
\item[(1)] Start with the initial state 
$\qubit{\psi^{(0)}_k} = \qubit{k}\qubit{0}\qubit{0}\qubit{0^{\ell}}$.

\item[(2)] Apply the quantum transform $\qubit{0}\rightarrow (1/\sqrt{M})\sum_{r\in I_n}\qubit{r}$ to the second register, and we 
obtain the superposition 
\[
\qubit{\psi^{(1)}_k} = \frac{1}{\sqrt{M}} \sum_{{r}\in I_n} \qubit{k}\qubit{{r}}\qubit{0}\qubit{0^{\ell}}.
\]

\item[(3)] Invoke a query to the oracle $\tilde{O}$ using the last three registers. The resulting quantum state is 
\[
\qubit{\psi^{(2)}_k} = \frac{1}{\sqrt{M}} \sum_{r \in I_n}\sum_{z\in \field_{q}}\alpha_{{r},z}
\qubit{k}\qubit{{r}}\qubit{z}\qubit{\phi_{{r},z}}.
\]

\item[(4)] To obtain a $k$-shuffled codeword state, we need to transform $\qubit{k}\qubit{r}\rightarrow \omega_{q}^{k\cdot r}\qubit{k}\qubit{r}$. This transform can be realized by the following series of simple transforms: $\qubit{k}\qubit{r}\qubit{0}\rightarrow \qubit{k}\qubit{r}\qubit{k\cdot r\;\mathrm{mod}\;q} \rightarrow \omega_{q}^{k\cdot r}\qubit{k}\qubit{r}\qubit{k\cdot r\;\mathrm{mod}\;q}\rightarrow \omega_{q}^{k\cdot r}\qubit{k}\qubit{r}\qubit{0}$. 
We then obtain the quantum state of the form 
\[
\qubit{\psi^{(3)}_k} =
 \frac{1}{\sqrt{M}} \sum_{r\in I_n} \sum_{z\in \field_{q}} \omega_{q}^{k\cdot z} \alpha_{{r},z} \qubit{k} \qubit{{r}}\qubit{z}\qubit{\phi_{{r},z}}.
\]
Since this step encodes the information on the first and the third resisters into the {\lq\lq}phase{\rq\rq}, for later convenience, this step will be referred to as {\em phase encoding}.

\item[(5)] Apply the inverse oracle $\tilde{O}^{-1}$ to the last three registers  and denote the resulting state  $(I \otimes \tilde{O}^{-1})\qubit{\psi^{(3)}_k}$ by $\qubit{\psi^{(4)}_{k}}$. When the oracle is called, 
the last three registers are changed in a unit time. 
The final state $\qubit{\psi^{(4)}_k}$ 
can be expressed in the form 
$
\kappa_{{x}}^{(k)}\qubit{k}\qubit{C_{{x}}^{(k)}}\qubit{\tau} + \qubit{{\Lambda}^{(k)}_{x}},
$ 
where $\qubit{\tau} = \qubit{0}\qubit{0^{\ell}}$ and  $(\bra{k}\bra{C_{{x}}^{(k)}}\bra{\tau})\qubit{{\Lambda}^{(k)}_{x}} =0$. 
\end{enumerate}
\vspace{1mm}
\hrule\vspace{5mm}

The execution time of $\AAA$ is clearly upper-bounded by a certain polynomial in $(n,\log_{2}q,\log_{2}M)$. Now, we want to calculate the amplitude $\kappa_{{x}}^{(k)}$. First, we note that 
\[
(I\otimes\tilde{O})\qubit{k}\qubit{C^{(k)}_{x}}\qubit{\tau} = \frac{1}{\sqrt{M}}\sum_{r\in I_n} \sum_{z\in \field_q} \omega_q^{k\cdot C_x(r)}\alpha_{r,z}\qubit{k}\qubit{r}\qubit{z}\qubit{\phi_{r,z}}. 
\]  
We denote the above state by the quantum state by $\qubit{\psi'}$.
Therefore, we have
\begin{eqnarray*}
\kappa^{(k)}_{x} &=& (\bra{k}\bra{C^{(k)}_{x}}\bra{\tau})((I\otimes\tilde{O}^{-1})\ket{\psi^{(3)}_{k}}) 
  \;\;=\;\; \braket{\psi'_{k,x}}{\psi^{(3)}_{k}} \\
&=& \frac{1}{M}\sum_{r\in I_n} \sum_{z\in\field_q} \omega_q^{k(z-C_x(r))}|\alpha_{r,z}|^2.
\end{eqnarray*}

The non-trivial part of our proof is to show a lower-bound of $|\kappa_{x}^{(k)}|$. Notice that a different proof appeared in \cite{KY06}.
By summing $\kappa_{x}^{(k)}$ over all $k\in\field_{q}^{+}$, the term $\sum_{k\in\field_{q}^{+}}|\kappa_{x}^{(k)}|$ is lower-bounded by 
\[
 \sum_{k\in\field^+_q} \left|\kappa_x^{(k)}\right|
\geq \left| \sum_{k\in\field^+_q} \frac{1}{M} \sum_{r\in I_n} \sum_{z\in \field_q} \omega_q^{k(z-C_x(r))}|\alpha_{r,z}|^2\right|
= \left| \sum_{k\in\field_{q}^{+}}\sum_{j\in\field_{q}}\omega_{q}^{k\cdot j}\left( \frac{1}{M}\sum_{r\in I_n}\left|\alpha_{r,C_x(r)+j}\right|^2  \right) \right|.
\]
We introduce a notation. For each value $j\in\field_q$, write $\beta_{j}$ for the term  
$(1/M)\sum_{r\in I_n} |\alpha_{r,C_x(r)+j}|^2$. Note that $\beta_0= \Pre_{\tilde{O}}(C_x)$ and $1-\beta_0 = \sum_{j\in\field_{q}^+}\beta_j$. Using this $\beta_j$-notation, we have
\begin{eqnarray*}
 \left| \sum_{k\in\field^+_q} \sum_{j\in \field_q} \omega_q^{k\cdot j}\beta_{j} \right| 
&=& \left| \sum_{k\in\field^+_q}\omega_q^{0} \beta_{0} + \sum_{k\in\field^+_q}\omega_q^{k} \beta_{1}+\cdots+\sum_{k\in\field^+_q}\omega_q^{(q-1)k} \beta_{q-1}\right|  \\
&=& \left| (q-1)\beta_0 - \sum_{j\in\field_q^+} \beta_{j} \right|\\
&=& \left| (q-1)\Pre_{\tilde{O}}(C_x) - (1-\Pre_{\tilde{O}}(C_x)) \right|\\
&=& \left| q\cdot \Pre_{\tilde{O}}(C_x) - 1\right|.
\end{eqnarray*}
Hence, we obtain $(1/(q-1))\sum_{k\in\field_{q}^{+}}|\kappa_{x}^{(k)}| \geq (1/(q-1))|q\cdot \Pre_{\tilde{O}}(C_x) - 1|$. This implies that there exists a number $k\in\field_q^+$ for which 
\[
 \left|\kappa_x^{(k)}\right|\geq 
\frac{1}{q-1}\left|q\cdot \Pre_{\tilde{O}}(C_x) - 1\right|
 = \frac{q}{q-1}\left| \Pre_{\tilde{O}}(C_x) - \frac{1}{q} \right|.
\]
This completes the proof.
\end{proofof}

Finally, we prove Theorem \ref{list-decoding}. 

\begin{proofof}{Theorem \ref{list-decoding}}
We assume that there exists a $(1-\nu)$-quantum codeword-state decoder $\DD$ for $C$ with $1-\nu(n)>\sqrt{1-\eta_{\varepsilon}(n)^2}$ for every number $n\in\nat$, where $\eta_{\varepsilon}(n) = (q(n)/(q(n)-1))\varepsilon(n)$. Fix $n\in\nat$.  Since $\DD$ is a $(1-\nu)$-quantum codeword-state decoder for $C$, for each $x\in\Sigma_n$ and $k\in\field_{q(n)}^{+}$, $\DD$ outputs $x$ from the $k$-shuffled codeword state $\ket{C_x^{(k)}}$ 
 with probability at least $1-\nu(n)$. Let $\tilde{O}$ be any quantumly corrupted codeword for $C$. 
Given $\tilde{O}$ as an implicit input, we consider the following algorithm $\BB$ that can solve the quantum list-decoding problem for $C$ with probability at least $\delta(n)$. Let $\AAA$ denote the quantum algorithm given in Lemma \ref{kappa}. 

For notational readability, we omit the script ``$n$.''  Write $\sigma$ for the value $1-\nu-\sqrt{1-\eta_{\varepsilon}^2}$. Initially, set $k=1$ in the algorithm $\BB$. 

\vspace{5mm}
\hrule\vspace{3mm}
\n{\sc Quantum Algorithm $\BB$:}
\begin{enumerate}\ms
\item[(1)] Starting with $\qubit{0^m}$, run the algorithm $\AAA$ to obtain the quantum state 
$\ket{\psi_k}$.

\item[(2)] Apply the algorithm $\DD$ to the second register of $\ket{\psi_k}$ using an appropriate number of ancilla qubits, say $m$. We then obtain the state
$\DD\ket{\psi_k}\ket{0^m}$.

\item[(3)] Measure the obtained state and 
add this measured result to the list of message candidates.

\item[(4)] Repeat Steps (1)--(3) $\ceilings{((1/\sigma)(\log_{e}{J_{\varepsilon,q,d,M}(n)}+\log_{e}(1/(1-\delta)))}$ 
times. 

\item[(5)] Repeat Steps (1)--(4) by incrementing $k$  by one at each repetition until $k=q-1$. Finally, output the list that is produced.
\end{enumerate}
\vspace{1mm}
\hrule\vspace{5mm}

For our convenience, we abbreviate as  $B_{\varepsilon}$ the set $\{x\in\Sigma_n\mid \Pre_{\tilde{O}}(C_x)\geq 1/q+\varepsilon\}$. Lemma \ref{kappa} implies that, for every element $x\in B_{\varepsilon}$, there exists an index $k\in\field_{q}^{+}$ such that $|\kappa_{x}^{(k)}|\geq \eta_{\varepsilon}$. By letting $X_{B_{\varepsilon}}^{(k)}= \{x\in B_{\varepsilon}\mid |\kappa_{x}^{(k)}|\geq \eta_{\varepsilon}\}$, we immediately obtain $B_{\varepsilon} = \bigcup_{k\in\field_{q}^{+}} X_{B_{\varepsilon}}^{(k)}$. We claim that the algorithm $\BB$ satisfies the following properties. 

\begin{claim}\label{closeness}
Let $k\in\field_{q}^{+}$.
\begin{enumerate}\vs{-2}
\item[(1)] Let $x\in X_{B_{\varepsilon}}^{(k)}$. With probability at least $\sigma$, we can observe $x$ when measuring the quantum state obtained after Step (2) in the computational basis.
\vs{-2}
\item[(2)] If we proceed Steps (1)--(3) 
$\ceilings{((1/\sigma)(\log_{e}{|X_{B_\varepsilon}^{(k)}|}+\log_{e}(1/(1-\delta)))}$ times for the same index $k$, then we obtain a list that includes all messages in  $X_{B_{\varepsilon}}^{(k)}$ with probability at least $\delta$.
\end{enumerate}
\end{claim}

Let us prove Claim \ref{closeness}. The {\em trace distance} $\|\rho-\sigma\|_{\mathrm{tr}}$ between two quantum states $\rho$ and $\sigma$ is  defined to be $Tr\sqrt{(\rho-\sigma)(\rho-\sigma)^{\dagger}}$. In particular, for two pure states $\qubit{\phi}$ and $\qubit{\psi}$, the trace distance between them can be calculated as  $\|\ketbra{\phi}{\phi}-\ketbra{\psi}{\psi}\|_{\mathrm{tr}} = 2\sqrt{1-|\braket{\phi}{\psi}|^2}$. For two (probability) distributions $D_1$ and $D_2$ over $\Sigma_n$, the $L_1$-norm (or the {\em total variation distance}) $\|D_1-D_2\|_1$ is defined as $\sum_{x\in\Sigma_n}|D_1(x)-D_2(x)|$. 

\begin{proofof}{Claim \ref{closeness}}
We fix $k\in\field_{q}^{+}$ arbitrarily.

(1) Choose any element $x\in X_{B_{\varepsilon}}^{(k)}$. Denote by $p_k(x)$ the probability of observing $x$ at Step (3) during round $k$. Our goal is to show that $p_k(x)\geq \sigma$. For simplicity, let $\qubit{\phi_{x,k}} = \qubit{k}\qubit{C^{(k)}_{x}}\qubit{\tau}\qubit{0^m}$ and $\qubit{\hat{\psi}_k} = \qubit{\psi_k}\qubit{0^m}$.
The trace distance between two pure states $\DD\ket{\hat{\psi}_{k}}$ 
and $\DD\ket{\phi_{x,k}}$ equals 
\[
\|\DD\ketbra{\phi_{x,k}}{\phi_{x,k}}\DD^{\dagger} - \DD\ketbra{\hat{\psi}_k}{\hat{\psi}_k}\DD^{\dagger}\|_{\mathrm{tr}} = 
\|\ketbra{\phi_{x,k}}{\phi_{x,k}} - \ketbra{\hat{\psi}_k}{\hat{\psi}_k}\|_{\mathrm{tr}} = 2\sqrt{1-|\braket{\phi_{x,k}}{\hat{\psi}_k}|^2} = 2\sqrt{1-|\kappa^{(k)}_{x}|^2}.
\]

Let $\tilde{D}_{k}(y)$ and $D_{x,k}(y)$ be the probabilities of obtaining $y\in\Sigma_n$ 
by measuring the states $\DD\ket{\hat{\psi}_k}$ and 
$\DD\ket{\phi_{x,k}}$, respectively, in the computational basis. 
Note that  $p_k(x)$ equals $\tilde{D}_{k}(x)$. 
Since the total variation distance between 
$\tilde{D}_{k}$ and $D_{x,k}$ is at most the trace distance between $\ket{k}\ket{C_x^{(k)}}\ket{\tau}$ and $\ket{\psi_k}$, 
it follows that 
\[
\|\tilde{D}_{k} - D_{x,k}\|_{1} \leq \|\DD\ketbra{\phi_{x,k}}{\phi_{x,k}}\DD^{\dagger} - \DD\ketbra{\hat{\psi}_k}{\hat{\psi}_k}\DD^{\dagger}\|_{\mathrm{tr}} = 2\sqrt{1-|\kappa^{(k)}_x|^2}.
\]
Moreover, we claim that $\|\tilde{D}_{k} - D_{x,k}\|_1\geq 2(1-\nu-\tilde{D}_{k}(x))$. This is shown as follows. First, we note that $\|\tilde{D}_{k}-\tilde{D}_{k,x}\|_1$ is lower-bounded by
\begin{eqnarray*}
 \|\tilde{D}_{k}-D_{x,k}\|_1 &=& |\tilde{D}_{k}(x)-D_{x,k}(x)| + \sum_{y:y\neq x} |\tilde{D}_{k}(y)-D_{x,k}(y)| \\
&\geq& |\tilde{D}_{k}(x) -D_{x,k}(x)| + \left|\sum_{y:y\neq x}\tilde{D}_{k}(y) - \sum_{y:y\neq x}D_{x,k}(y)\right| = 2|\tilde{D}_{k}(x) - D_{x,k}(x)|
\end{eqnarray*}
since $\sum_{y\in\Sigma_n}\tilde{D}_{k}(y)=\sum_{y\in\Sigma_n}D_{x,k}(y)=1$. 
We then obtain
\[
\|\tilde{D}_{k} - D_{x,k}\|_1  \geq 2\left| \tilde{D}_{k}(x)-\tilde{D}_{k,x}(x) \right| \geq 2(1-\nu- \tilde{D}_{k}(x)).
\]
The above two bounds on $\|\tilde{D}_{k}-\tilde{D}_{k,x}\|_1$ yields the following inequality: 
\[
 1-\nu- \tilde{D}_{k}(x) \leq \sqrt{1-|\kappa^{(k)}_{x}|^2},
\]
which immediately implies 
\[
 \tilde{D}_{k}(x) \geq 1-\nu-\sqrt{1-|\kappa^{(k)}_{x}|^2} \geq 1 - \nu - \sqrt{1-\eta_{\varepsilon}^2} = \sigma
\]
since $|\kappa^{(k)}_{x}|\geq \eta_{\varepsilon}$. Therefore, we conclude that $p_k(x) \geq \sigma$, as requested. 

\ms

(2) Assuming that Steps (1)--(3) for the same index $k$ are repeated $t$ times to create a list including all the elements in $X_{B_{\varepsilon}}^{(k)}$, we wish to prove that $t\geq  ((1/\sigma)(\log_{e}{|X_{B_\varepsilon}^{(k)}|}+\log_{e}{(1/(1-\delta))})$. This gives the desired bound since $|X_{B_{\varepsilon}}^{(k)}|\leq J_{\varepsilon,q,d,M}(n)$. 
Note that we obtain $x\in X_{B_\varepsilon}^{(k)}$ through these steps 
 with probability at least $\sigma$. This implies that, for each fixed element $x_0\in X_{B_\varepsilon}^{(k)}$, the probability of obtaining no $x_0$  
within $t$ samples is upper-bounded by $(1-\sigma)^t$. Therefore, with probability at most $|X_{B_\varepsilon}^{(k)}|(1-\sigma)^t$, there exists an $x\in X_{B_\varepsilon}^{(k)}$ for which $t$ samples does not contain $x$. 

Since the probability of obtaining the desired valid list is at least $\delta$, we demand the condition that $|X_{B_{\varepsilon}}^{(k)}|(1-\sigma)^t\leq 1-\delta$; equivalently, 
\[
t\log_{e}\frac{1}{1-\sigma}\geq \log_{e}|X_{B_{\varepsilon}}^{(k)}| + \log_{e}\frac{1}{1-\delta},
\]
which yields the desired bound
\[
t\geq \frac{1}{\sigma}\left(\log_{e}{|X_{B_\varepsilon}^{(k)}|}+\log_{e}{\frac{1}{1-\delta}}\right)
\]
because $\log_{e}(1/(1-\sigma))$ is lower-bounded by 
\[
\log_{e}\frac{1}{1-\sigma}= - \log_{e}(1 - \sigma) 
= \sum_{i=1}^{\infty}\frac{\sigma^i}{i} \geq \sigma.
\] 
This completes the proof of the claim.
\end{proofof}

Claim \ref{closeness} guarantees that, since $|X_{B_{\varepsilon}}^{(k)}|\leq |B_{\varepsilon}| \leq J_{\varepsilon,q,d,M}(n)$, for all indices $k\in\field_{q}^{+}$, if we run Steps (1)-(4), then we obtain a list containing all the elements in $B_{\varepsilon}$ with probability at least $\delta$. Note that, at Step (3), we add only one element into our list of candidates. Hence, the size of the list generated by $\BB$ is at most  
$q(n)-1$ times $\ceilings{((1/\sigma)(\log_{e}{|B_\varepsilon|}+\log_{e}{(1/(1-\delta))})}$. 
It is obvious that the total number of queries is at most twice as many as 
the list size. 
\end{proofof}

\section{Nearly Phase-Orthogonal Codes}\label{sec:phase-orthogonal}

Through Sections \ref{sec:hardcore-property}--\ref{sec:list-decoding}, we have developed a solid foundation for the proof of our main theorem, presented in Section \ref{sec:new-hardcore}, concerning the quantum hardcore property of three classical block codes. In this section, we shall target two codes, $q$-ary Hadamard codes and pairwise equality codes, which share a common feature, called {\em nearly phase-orthogonality}. The proof of their quantum hardcore property will exploit this feature. 

\subsection{Nearly Phase-Orthogonality}

Now, we want to introduce a family of {\em nearly phase-orthogonal codes}.
Let us first recall that,
for any $(M(n),n)_{q(n)}$-code family with a message space $\Sigma^*=\bigcup_{n\in\nat}\Sigma_n$ and a series $\{I_n\}_{n\in\nat}$ of index sets, a {\em $k$-shuffled codeword state} for a codeword $C_{x}$ that encodes a message ${x}\in \Sigma_n$ is the quantum state 
$
\qubit{C_{x}^{(k)}} = \frac{1}{\sqrt{M(n)}}\sum_{{r}\in I_{n}}\omega_{q(n)}^{k\cdot C_{{x}}({r})}\qubit{{r}}. 
$ 

\begin{definition}[nearly phase orthogonality]
Let $\eta$ be any function from $\nat$ to the unit real interval $[0,1]$. We say that a classical block $(M(n),n)_{q(n)}$-code family $C$ with a message space $\Sigma=\bigcup_{n\in\nat}\Sigma_n$ is said to be {\em $\eta$-nearly phase-orthogonal} if $|\braket{C_x^{(k)}}{C_y^{(k)}}|\leq\eta(n)$ for any number $n\in\nat$, any element $k\in\field_{q(n)}^{+}$, and any message pair $x,y\in \Sigma_n$ with $x\neq y$.
In particular, any $0$-nearly phase-orthogonal code is called {\em phase orthogonal}.
\end{definition}

Notice that, since $k$-shuffled codeword states are pure quantum states, the value  $|\braket{C_{{x}}^{(k)}}{C_{{y}}^{(k)}}|$ coincides with the {\em fidelity} $F(\qubit{C_{x}^{(k)}},\qubit{C_y^{(k)}})$ of $\qubit{C_x^{(k)}}$ and $\qubit{C_y^{(k)}}$. 

The Hadamard code  $\mathrm{Had}^{(q)}$  is an example of phase-orthogonal code and its phase orthogonality plays an important role in the proof of the quantum hardcore property of $\HAD^{(q)}$ in Section \ref{sec:theorem-2.3}.
In general, the phase-orthogonality of an $(M(n),n)_{q(n)}$-code family $C$ with a message space $\Sigma^*=\bigcup_{n\in\nat}\Sigma_n$ implies that, for each pair $(n,k)\in \nat\times \field_{q(n)}^{+}$, the set  $\{\qubit{C^{(k)}_0},\qubit{C^{(k)}_1},\ldots,\qubit{C^{(k)}_{N(n)-1}}\}$ forms an orthonormal basis of an $N(n)$-dimensional Hilbert space, where $N(n)=|\Sigma_n|$. 
Note that, since a quantum hardcore $C(x,r)$ requires the condition that $|r|=poly(|x|)$, it suffices to consider only $(M(n),n)_{q(n)}$-codes  $C$ that satisfy the inequality $M(n)\geq N(n)$. 
For any binary code, in particular, its phase-orthogonality can be  naturally induced from the standard {\em inner product} of two codewords when we translate their binary symbols $\{0,1\}$ into $\{+1,-1\}$.

An immediate benefit of the phase-orthogonality is explained as follows. If a code $C=\{C_{{x}}\}_{{x}\in\Sigma^*}$ is phase-orthogonal, then we can strengthen Lemma \ref{kappa} so that we can isolate simultaneously all individual messages $x$. 
Moreover, a nearly phase-orthogonal code $C$ can provide a lower bound on the (Hamming) distance of $C$. 

\begin{proposition}\label{distance-bound}
Let $C=\{C_x\}_{x\in\Sigma^*}$ be any $(M(n),n,d(n))_{q(n)}$-code. Let $\eta$ be any function such that $0\leq \eta(n)\leq1$ for any number $n\in\nat$.
If $C$ is $\eta$-nearly phase-orthogonal, the distance $d(n)$ is lower-bounded by $(1-\eta(n))M(n)/2$ for every length $n$.
\end{proposition}

Proposition \ref{distance-bound} can be obtained from the following lemma, which relates the fidelity $F(\qubit{C_{x}^{(k)}},\qubit{C_y^{(k)}})$ to the relative Hamming distance $\Delta(C_x,C_y)$. The proof of the lemma is found in Appendix for readability. 

\begin{lemma}\label{fidelity}
For any pair $(C_x,C_y)$ of codewords in a given $(M(n),n,d(n))_{q(n)}$-code $C$ and for any index $k\in\field_q^{+}$, 
\[
F(\qubit{C^{(k)}_x},\qubit{C^{(k)}_y}) \geq 1- 2\Delta(C_x,C_y),
\]
where the equality holds for any binary code $C$.
\end{lemma}

Let us explain how to prove Proposition \ref{distance-bound} from Lemma \ref{fidelity}. Let $C$ be any $(M(n),n,d(n))_{q(n)}$-code that is $\eta$-nearly phase-orthogonal. {}From this nearly phase-orthogonality, the function $\eta$ satisfies that $\eta(n)\geq |\braket{C^{(k)}_x}{C^{(k)}_y}|$ for all parameters $k,x,y$ with $x\neq y$. Apply Lemma \ref{fidelity} and we obtain $\eta(n)\geq 1-2d(C_x,C_y)/M(n)$, from which we conclude that $d(C_x,C_y)\geq (1-\eta(n))M(n)/2$. Because $d(n)$ is the distance, it follows that $d(n)\geq d(C_x,C_y)\geq (1-\eta(n))M(n)/2$. This completes the proof of Proposition \ref{distance-bound}.

\subsection{Proof of Theorem \ref{new-hardcore}(1-2)}\label{sec:theorem-2.3}

How can we prove Theorem \ref{new-hardcore}? As we have discussed in Sections \ref{sec:hardcore-property}--\ref{sec:list-decoding}, with help of Theorem \ref{code-hardcore} and Corollary \ref{negligible-code}, we can prove the quantum hardcore property of each function, given in Theorem \ref{new-hardcore}, by constructing its polynomial-time $(1-s)$-quantum codeword-state decoder with a certain negligible function $s$. 
In this subsection, we target two code families: $q$-ary Hadamard codes $\HAD^{(q)}$ and pairwise equality codes $\PEQ$. Our quantum codeword-state decoders for these codes are obtained by exploiting their nearly phase-orthogonality.

For our proof, we assume the following limited form of {\em quantum Fourier transform} $F_n$, over a finite additive group $\integer_{n}$, running in time polynomial in $n$:  for any element $s\in\integer_n$, 
\[
F_{n}:\qubit{s}\rightarrow \frac{1}{\sqrt{n}} \sum_{r\in \integer_n}\omega_n^{s\cdot r}\qubit{r}.
\]  
For a more general form of quantum Fourier transform over a finite field $\field_{n}$, see, \eg \cite{DHI03}, where the following statement is proven: for any prime power $q$, $F_{q}$ can be approximated to within error $\varepsilon$ in time polynomial in $(\log_{2}{q},\log_{2}(1/\varepsilon))$.

\begin{proofof}{Theorem \ref{new-hardcore}(1-2)}
It suffices to provide a negligible function $s$ and a polynomial-time $(1-s)$-quantum codeword-state decoder for each of the given codewords $C$ in Theorem \ref{new-hardcore}(1-2). {}From such a decoder, by Theorem \ref{negligible-code}, we can construct a polynomial-time quantum list-decoder for $C$. 
Theorem \ref{code-hardcore} then guarantees the quantum hardcore property of $C$.
In our case of $\HAD^{(q)}$ and $\PEQ$, we can utilize their phase-orthogonality and build a polynomial-time $(1-2^{-n})$-quantum codeword-state decoder for them.

Now, fix $n\in\nat$ and omit ``$n$'' for readability. 

\ms

(1) The simple case $q=2$ was implicitly proven by Adcock and Cleve \cite{AC02} and also by Bernstein and Vazirani \cite{BV97}. 
Consider the general case $q\geq2$. Obviously, $\HAD^{(q)}$ is polynomial-time computable. Now, we intend to show that $\HAD^{(q)}$ is $(1-2^{-n})$-quantum codeword-state decodable. 
Let  $\qubit{\HAD^{(q)}}$ be a codeword state. Consider the quantum 
Fourier transform $F_q$ over $\field_{q}$.   
To recover $x$ from the codeword state $\qubit{\HAD^{(q)}}$, we note that 
\[
F_q\qubit{x}=\frac{1}{\sqrt{q}}\sum_{r\in\field_q}\omega_q^{x\cdot r}\qubit{r} = \frac{1}{\sqrt{q}}\sum_{r\in\field_q}\omega_q^{\HAD^{(q)}_x(r)}\qubit{r} = \qubit{\HAD^{(q)}_x}.
\]
Hence, apply the inverse of $F_{q}$ to $\qubit{\HAD^{(q)}_x}$ and we immediately obtain $\qubit{x}$ with probability $1$. Since $q$ is an arbitrary prime number, we may not simulate $F_{q}$ exactly. Instead of applying $F_{q}$, however, we can use its approximation whose approximation error is exponentially small. 
Therefore, we conclude that $\HAD^{(q)}$ is indeed $(1-2^{-n})$-quantum codeword-state decodable, as requested.

\ms

(2) We want to prove that $\PEQ$ has a polynomial-time $(1-2^{-n})$-quantum codeword-state decoder. We first observe the following key equation: 
\begin{eqnarray*}
\qubit{\PEQ_x(r)} &=& \frac{1}{\sqrt{2^n}} \sum_{r=0}(-1)^{\PEQ_x(r)}\ket{r} \\
&=& 
\frac{1}{\sqrt{4}} \sum_{r_1,r_2} (-1)^{\EQ(x_1x_2,r_1r_2)} \ket{r_1r_2} \otimes \cdots \otimes \frac{1}{\sqrt{4}}\sum_{r_{n-1},r_n}(-1)^{\EQ(x_{n-1}x_n,r_{n-1}r_n)}\ket{r_{n-1}r_n}.
\end{eqnarray*}
Let us consider the following unitary transform $H_C$, which we call the {\em circulant Hadamard transform}:  
\[
 H_C =_{def} \;\; 
{\small \left(\begin{array}{cccc}
-1 & 1 & 1 & 1\\
1 & -1 & 1 & 1\\
1 & 1 & -1 & 1\\
1 & 1 & 1 & -1
\end{array}\right) } \;\;=\;\;
F_4^{-1}
{\small \left(\begin{array}{cccc}
-1 & 0 & 0 & 0\\
0 & -1 & 0 & 0\\
0 & 0 & -1 & 0\\
0 & 0 & 0 & -1
\end{array}\right) } F_4,
\]
where $F_4$ is the quantum Fourier transform over $\field_4$. 
Since $H_{C}$ satisfies the equality 
\[
 H_C\left(\frac{1}{\sqrt{4}} \sum_{r_i,r_{i+1}}(-1)^{{\rm EQ}(x_ix_{i+1},r_ir_{i+1})}\ket{r_ir_{i+1}}\right) = \ket{x_ix_{i+1}},
\]
we can obtain $\qubit{\phi}= \qubit{x_1x_2}\otimes\cdots\otimes\qubit{x_{n-1}x_n}$ from the codeword state $\ket{\PEQ_x}$ by applying $U=H_C^{\otimes n/2}$. 
Using an approximation of $F_4$, we can simulate the unitary operator $H_{C}$ with exponentially small error. Finally, from the quantum state $\qubit{\phi}$, 
we can easily deduce $x$. Therefore, we obtain a polynomial-time $(1-2^{-n})$-quantum codeword-state decoder for $\PEQ$. 
\end{proofof}

\subsection{Query Complexity of Quantum List-Decoders}\label{sec:theorem-2.6}

Whereas any known classical list-decoder for a $q$-ary Hadamard code needs a polynomial number of queries per each message candidate, our quantum list-decoder constructed in the previous subsection requires only two queries. For more general nearly phase-orthogonal codes, we shall show that these codes have quantum list-decoders that make a significantly small number of queries to a given quantumly corrupted codeword.  

\begin{theorem}\label{phase-orthogonal-circuit}
Let $C = \{(\Sigma_n,I_n,\Gamma_n)\}_{n\in\nat}$ be any $(M(n),n)_{q(n)}$-code family and assume that $q(n)\in n^{O(1)}$ and $\log_{2}M(n)\in n^{O(1)}$.  Let $\eta$ be any function from $\nat$ to the unit real interval $[0,1]$  and assume that $\eta(n)|\Sigma_n|$ is a 
negligible 
function. If $C$ is $\eta$-nearly phase-orthogonal, then there exists a quantum list-decoder for $C$ whose query complexity is propotional to the size of its output list.  
\end{theorem}


The proof of Theorem \ref{phase-orthogonal-circuit} requires the following technical statement on nearly phase-orthogonal codes. We show that certain types of nearly phase-orthogonal codes are indeed quantum list-decodable with low query complexity. 

\begin{proposition}\label{codeword-to-listdecode}
\sloppy Let $C$ be any $(M(n),n)_{q(n)}$-code family with a message space $\Sigma^*=\bigcup_{n\in\nat}\Sigma_n$. Let $N(n)=|\Sigma_n|$ for each length $n$. Let $\varepsilon$, $\delta$, and $\eta$ be any three functions from $\nat$ to the real interval $[0,1]$, and also let $\eta'(n)=\eta(n)N(n)$ and $\eta_{\varepsilon}(n)=(q(n)/(q(n)-1))\varepsilon(n)$ for every number $n\in
\nat$. Let $\ell$ be any polynomially-bounded function from $\nat$ to $\nat$. 
Assume that $\eta'$ is a negligible function. 
Assume that there is a positive-valued function $\sigma$ for which its reciprocal is upper-bounded by a certain positive polynomial in $(n,q(n),\log_{2}M(n),1/\varepsilon(n),\log_{2}(1/(1-\delta(n))))$ and it also satisfies $1-\eta'(n) - \sigma(n) > \sqrt{1-\eta_{\varepsilon}(n)^2}$ for every number $n\in\nat$.  If $C$ is $\eta$-nearly phase-orthogonal, then $C$ is $(\ell,\varepsilon,\delta)$-quantum list-decodable with list size polynomial in $(n,q(n),\log_{2}M(n),1/\varepsilon(n),\log_{2}(1/(1-\delta(n))))$ and query complexity is at most twice as many as the list size.
\end{proposition}

We delay the proof of this proposition and quickly prove Theorem \ref{phase-orthogonal-circuit}. With help of the proposition, 
the proof of this theorem is similar to that of Theorem \ref{code-hardcore}. 

\begin{proofof}{Theorem \ref{phase-orthogonal-circuit}}
Let $C=\{(\Sigma_n,I_n,\Gamma_n)\}_{n\in\nat}$ be any $(M(n),n)_{q(n)}$-code family. Write $N(n)=|\Sigma_n|$.   
For any number $n\in\nat^{+}$, define $\delta(n)=1-2^{-n}$, which implies $\log_{2}(1/(1-\delta(n)))=n$.
Let $\varepsilon$ be any noticeable function from $\nat$ to $[0,1]$. Take a positive polynomial $\hat{p}$ and assume that $\varepsilon(n)\geq 1/\hat{p}(n)$ for all numbers $n\in\nat$. Let $\eta$ be any function such that $\eta'(n)=\eta(n)N(n)$ is negligible. Since $\eta'$ is negligible, it follows that, for all sufficiently large numbers $n$, $\eta'(n)$ is upper-bounded by $1/4\hat{p}(n)^2$.  

Let us define $\sigma$ as $\sigma(n)= 1-\eta'(n) - \sqrt{1-\eta_{\varepsilon}(n)^2}$, where $\eta_{\varepsilon}(n)=(q(n)/(q(n)-1))\varepsilon(n)$. Similar to the proof of Corollary \ref{negligible-code}, 
the function $1/\sigma(n)$ is upper-bounded by $4\hat{p}(n)^2$. Since $\log_{2}M(n)\in n^{O(1)}$ and $q(n)\in n^{O(1)}$, we can conclude that 
$1/\sigma(n)$ is bounded from above by a certain polynomial in $(n,q(n),\log_{2}M(n),1/\varepsilon(n),\log_{2}(1/(1-\delta(n))))$ . 
Now, apply Proposition \ref{codeword-to-listdecode} and we obtain an $(\ell,\varepsilon,\delta)$-quantum list-decoder $\DD$ for $C$ with list size polynomial in $n$.
\end{proofof}

In the rest of this subsection, we shall give the proof of Proposition \ref{codeword-to-listdecode}.
This proposition follows directly from Theorem \ref{list-decoding} and the next key lemma, 
which  states  that any nearly phase-orthogonal code has a certain type of quantum codeword-state decoder. 

\begin{lemma}\label{almost-orthogonal-code}
Let $C$ be any $(M(n),n)_{q(n)}$-code family with a message space $\Sigma^*=\bigcup_{n\in\nat}\Sigma_n$ such that $M(n)\geq N(n)$ for all numbers  $n\in\nat$, where $N(n)=|\Sigma_n|$. Let $\eta$ be any function from $\nat$ such that $0\leq \eta(n)N(n) \leq 1$ for any sufficiently large number $n\in\nat$. If $C$ is $\eta$-nearly phase-orthogonal, then there exists a $(1-\eta(n)N(n))$-quantum codeword-state decoder for $C$. 
\end{lemma}

\sloppy Lemma \ref{almost-orthogonal-code} helps us prove Proposition \ref{codeword-to-listdecode} in the following fashion. Assuming the premise of the proposition, take $\eta'(n)$ and take $\sigma$. Note that $1/\sigma(n)$ is upper-bounded by a certain polynomial in $(n,q(n),\log_{2}M(n),1/\varepsilon(n),\log_{2}(1/(1-\delta(n))))$.  
For an given $\eta$-nearly phase-orthogonal code $C$, by Lemma \ref{almost-orthogonal-code}, we obtain a $(1-\eta'(n))$-quantum codeword-state decoder. Theorem \ref{list-decoding} then guarantees the existence of an $(\ell,\varepsilon,\delta)$-quantum list-decoder for $C$ whose list size is polynomial in $(n,q(n),\log_{2}M(n),1/\varepsilon(n),\log_{2}(1/(1-\delta(n))))$. This list-decoder has query complexity of at most twice as many as the list size. This completes the proof of the proposition.

\ms

To prove Lemma \ref{almost-orthogonal-code}, we use the notion of pretty-good measurement (known also as square-root measurement or least-squares measurement) \cite{EF01,HW94}. 
Let $E_n$ denote the $n$-by-$n$ identity matrix.
Before the proof, we note the following lemma, which can be obtained by following the proof of \cite[Lemma~9.1]{Alo03}, in which the case of real matrices is considered. Here, we treat every quantum state $\qubit{\phi}$ as a column vector.

\begin{lemma}\label{Alon-lemma}
Let $\eta\in[0,1]$. Let $G=(\eta_{i,j})_{i,j\in[N]}$ be any complex symmetric $N$-by-$N$ matrix with
$\eta_{i,i}=1$ and $|\eta_{i,j}|\le\varepsilon$ for all pairs $(i,j)\in[N]^2$ with $i\neq j$. It then holds that 
\[
\mathrm{rank}(G) \ge \frac{N}{1+(N-1)\varepsilon^2}.
\]
\end{lemma}

\begin{proofof}{Lemma \ref{almost-orthogonal-code}}
Fix $n\in\nat$ and $k\in\field_{q(n)}^{+}$. For simplicity, let   $I_N=[0,N(n)-1]_{\integer}$. For readability, we omit the script ``$n$'' in the rest of this proof.  We wish to construct a quantum algorithm $\AAA$  whose success probability of obtaining $z$ from $\ket{C^{(k)}_z}$ is 
at least $1-\eta N$ whenever $|\braket{C^{(k)}_x}{C^{(k)}_y}|\leq \eta$ 
for any two distinct messages $x,y\in\Sigma_n$.

We want to design $\AAA$ by following an argument of pretty good measurement \cite{EF01,HW94}. Let $S$ be the $M$-by-$N$ matrix $(\qubit{C^{(k)}_0},\qubit{C^{(k)}_1},\ldots,\qubit{C^{(k)}_{N-1}})$, in which the $i$th column of $S$ expresses the column vector $\qubit{C^{(k)}_i}$. Notice that $S\qubit{z}_{N}=\qubit{C^{(k)}_z}$ for each $z\in I_{N}$, where $\qubit{z}_{N}$ denotes the $N$-dimensional unit vector whose $z$th entry is $1$ and $0$ elsewhere.

Note that $S^{\dag}S$ is an $N$-by-$N$ matrix. We first show that $\mathrm{rank}(S^{\dag}S)=N$, which implies that all eigenvalues of $S^{\dag}S$ are non-zero.
Setting $G = S^{\dag}S$ and $\varepsilon = \eta$, Lemma \ref{Alon-lemma} gives
\[
\mathrm{rank}(S^{\dag}S) 
\ge \frac{N}{1+(N-1)\eta^2} > \left(1-(N-1)\eta^2\right)N \ge N - 1 + \frac{1}{N}.
\]
We used the facts that $1/(1+\delta) > 1-\delta$ for any $\delta < 1$ 
at the second inequality and $\eta \le 1/N$ at the last inequality.
Since the rank is an integer larger than $N-1+(1/N)$, it should be exactly $N$, concluding that $\mathrm{rank}(S^{\dag}S)=N$.

Since $S^{\dag}S$ is Hermitian and positive definite, it has a set of positive eigenvalues, say 
$\{\lambda_0,\ldots,\lambda_{N-1}\}$. Let $\lambda_{\rm min}= \min\{\lambda_0,\lambda_1,\ldots,\lambda_{N-1} \}>0$.
We claim that $\lambda_{\rm min}$ is relatively large.

\begin{claim}\label{lambda-bound}
$\lambda_{\rm min}\geq 1- \eta N$.
\end{claim}

\begin{proof}
Let $G=S^{\dag}S$, the $N$-by-$N$ matrix $(\eta_{i,j})_{i,j}$,
where $\eta_{i,j} = \braket{C^{(k)}_i}{C^{(k)}_j}$. By our assumption, it follows that $|\eta_{ij}|\leq \eta$ for any pairs $(i,j)$.
Since $G$ is Hermitian, 
let $G=\sum_{i=0}^{N-1} \lambda_i\ketbra{\psi_i}{\psi_i}$ be a spectral decomposition of $G$ for the eigenstates $\{\ket{\psi_i}\}_{i\in I_N}$. We then have
\[
\min_{\norm{\ket{\psi}}=1}|\bra{\psi}G\ket{\psi}| = 
\min_{\norm{\ket{\psi}}=1}\left|\sum_{i=0}^{N-1} 
\lambda_i|\braket{\psi_i}{\psi}|^2\right| =\lambda_{\rm min}.
\]
Note that, for any state $\qubit{\psi}$ of the form $\sum_{i\in I_{N}} \alpha_i\ket{i}$ with complex numbers $\alpha_i$'s, the value $|\bra{\psi}G\ket{\psi}|$ equals
\[
|\bra{\psi}G\ket{\psi}|
= \left|1+\sum_{i\neq j}\eta_{i,j}\alpha^*_i\alpha_j\right|
= \left|1 -\eta \sum_{i<j}(|\alpha_i|^2 + |\alpha_j|^2) 
   + \sum_{i<j} \eta|\alpha_i|^2 + \eta_{i,j}\alpha_i^*\alpha_j + \eta_{j,i}\alpha_j^*\alpha_i + \eta|\alpha_j|^2\right|.
\]
We then focus on the term $\sum_{i<j}\eta|\alpha_i|^2 + \eta_{i,j}\alpha_i^*\alpha_j + \eta_{j,i}\alpha_j^*\alpha_i + \eta|\alpha_j|^2$. We will show that the term is real positive. Let $\gamma_{i,j} = \eta_{i,j}/|\eta_{i,j}|$. 
Since $\eta_{j,i}=\eta_{i,j}^*$ and $\eta-|\eta_{i,j}|\ge 0$, we have
\begin{eqnarray*}
\lefteqn{\sum_{i<j} \eta|\alpha_i|^2 + \eta_{i,j}\alpha_i^*\alpha_j + \eta_{j,i}\alpha_j^*\alpha_i + \eta|\alpha_j|^2}\hs{10} \\
&=& 
\sum_{i<j} (\eta-|\eta_{i,j}|)|\alpha_i|^2 + (\eta-|\eta_{i,j}|)|\alpha_j|^2
+ |\eta_{i,j}|(|\alpha_i|^2 + \gamma_{i,j}\alpha_i^*\alpha_j + \gamma_{i,j}^*\alpha_j^*\alpha_i + |\alpha_j|^2)\\
&=& 
\sum_{i<j} (\eta-|\eta_{i,j}|)|\alpha_i|^2 + (\eta-|\eta_{i,j}|)|\alpha_j|^2
+ |\eta_{i,j}|(\gamma_{i,j}^*\alpha_i + \alpha_j)(\gamma_{i,j}\alpha_i^*+\alpha_j^*)\\
&=& 
\sum_{i<j} (\eta-|\eta_{i,j}|)|\alpha_i|^2 + (\eta-|\eta_{i,j}|)|\alpha_j|^2
+ |\eta_{i,j}||\gamma_{i,j}^*\alpha_i + \alpha_j|^2 \ge 0.
\end{eqnarray*}
Therefore, we have 
\[
\min_{\norm{\ket{\psi}}=1}|\bra{\psi}G\ket{\psi}| 
\ge \left|1 -\eta \sum_{i<j}(|\alpha_i|^2 + |\alpha_j|^2)\right|
\ge 1 -\eta N.
\]
\end{proof}

We continue our proof of Lemma \ref{almost-orthogonal-code}. Let $S=PTQ^{\dag}$ be a {\em singular-value decomposition} (see, \eg \cite{HJ85}), where  
$P$ is an $M$-by-$M$ unitary matrix, $Q$ is an $N$-by-$N$ unitary matrix, and $T$ is an $M$-by-$N$ matrix of the form $\tinycomb{T'}{O}$ with the diagonal matrix $T'={\rm diag}(\sqrt{\lambda_0}, \sqrt{\lambda_1},\ldots,\sqrt{\lambda_{N-1}})$.
We therefore have $\bra{z}_M US\ket{z}_N=\bra{z}_M UPTQ^{\dag}\ket{z}_N$ for any $z\in I_N$.

We define an $M$-by-$M$ matrix $U$ as $U=RP^\dag$, where the $M$-by-$M$ matrix $R$ is
\[
R = 
\left( 
\begin{array}{cc} 
Q    & O \\
O & E_{M-N}
\end{array}
\right).
\]
It immediately follows that, for any $z\in I_N$, 
\begin{eqnarray*} 
\bra{z}_M US\ket{z}_N &=& \bra{z}_M RTQ^{\dag}\ket{z}_N 
\;\;=\;\; \bra{z}_M \smatrix{Q}{O}{O}{E_{M-N}} \comb{T'}{O} Q^{\dagger} \ket{z}_N \\
&=& \bra{z}_M  \comb{QT'Q^{\dagger}}{O} \ket{z}_N 
\;\;=\;\; \bra{z}_N Q T'Q^{\dag} \ket{z}_N.
\end{eqnarray*}
The desired quantum algorithm $\AAA$ applies $U$ and then measures its result. The above calculation indicates that the probability of $\AAA$'s recovering $z$ from $\ket{C_z}$ is therefore lower-bounded by  
$|\bra{z}_{N}Q T' Q^{\dag} \ket{z}_{N} |^2$, which is at least $\lambda_{\rm min}$. The above claim yields the desired conclusion. 
\end{proofof}

\section{Circulant Codes}\label{sec:circulant}

To design quantum codeword-state decoders, Proposition \ref{codeword-to-listdecode} gives a constructive but not time-efficint method for nearly phase-orthogonal codes. Under a certain condition, we can construct quantum codeword-state decoders that run in {\em polynomial time}. One of such conditions is ``circulantness'' of codes with a certain property.  A {\em circulant code} family $C=\{C_i\}_{i\in\nat}$ requires its associate matrices $M^{(n)}_{C}=(C_i(j))_{i,j\in\nat}$ to be ``circulant.'' An example of such  code families is the shifted Legendre symbol code $\mathrm{SLS}^{p}$ described in Theorem \ref{new-hardcore}. Earlier, van Dam, Hallgren, and Ip \cite{DHI03} discussed, in essence, the quantum codeword-state decoding of $\SLS^{p}$ in the context of hidden shift  problems. With our notion of $k$-shuffled codeword states, we take a general approach toward circulant codes and give a broader insight into their quantum list-decodability.

\subsection{Circulantness and Fourier Transforms}

We formally introduce the notion of circulant codes. First, we fix a positive integer $n$ and let $L_n=[0,n-1]_{\integer}$. An $n$-by-$n$ integer matrix $Q=(q_{ij})_{i,j\in[n]}$ is called the {\em cyclic permutation matrix} if $q_{n,1}=1$, $q_{i,i+1}=1$ for any  index $i\in[n-1]$, and the others are all zeros. Notice that $Q^n$ equals the identity matrix. A {\em circulant matrix} $M$ is of the form $\sum_{j\in L_n}a_jQ^{j}$ for certain complex numbers $\{a_j\}_{j\in L_n}$; in other words, the $(i,j)$-entry of $M$ is $a_{j-i\;\mathrm{mod}\;n}$ for each pair $i,j\in L_n$. 

\begin{definition}[circulant code]\label{circulant-def}
A classical block code family $C=\{C_i\}_{i\in\nat}$ with index sets $\{I_n\}_{n\in\nat}$ is said to be {\em circulant}\footnote{This notion is different from the codes that have {\em circulant constructions} (see, \eg \cite{HP03}).} if, for every message length $n\in\nat$, the matrix $M^{(n)}_C= (C_i(j))_{i,j\in I_n}$ is circulant; namely, $M^{(n)}_C=\sum_{i=0}^{|I_n|-1}C_0(i)Q^i$, where $Q$ denotes the $M$-by-$M$ cyclic permutation matrix. 
\end{definition}

Circulant codes are desirable candidates for quantum hardcore functions. We shall give the proof of Theorem \ref{new-hardcore}(3) by exploiting the circulantness of $\SLS^{p}$.

The (discrete and quantum) Fourier transform is one of the most useful operations in use. Notice that circulant matrices can be diagonalized by these Fourier transforms. 
Consider the quantum Fourier transform $F_n$ over $\integer_{n}$. Any circulant matrix $M=\sum_{j\in L_n}a_jQ^j$ can be diagonalized by $F_n$ as follows:
\[
F_n^{-1}MF_n = \mathrm{diag}\left( \sum_{j\in L_n}a_j\omega_n^{i\cdot j}\right)_{i\in L_n} = \mathrm{diag}\left(\sum_{j\in L_n}a_j,\sum_{j\in L_n}a_j\omega_n^{j},\sum_{j\in L_n}a_j\omega_n^{2j}\ldots,\sum_{j\in L_n}a_j\omega_n^{(n-1)j}\right). 
\]

{}From the definition of $M_{C}^{(n)}$ in Definition \ref{circulant-def}, the transposed matrix $(M_{C}^{(n)})^t =(C_{j}(i))_{i,j\in I_n}$ can be expressed as $\sum_{j=0}^{|I_n|-1}C_0(|I_n|-j\;\mathrm{mod}\,|I_n|)Q^j$ and therefore it is also a circulant matrix.

Now, we focus our attention on {\em $k$-shuffled codeword states} of circulant codes. Let $C$ be any $(M(n),n)_{q(n)}$ circulant code with a series $\{I_n\}_{n\in\nat}$ of index sets. Consider $k$-shuffled codeword states $\qubit{C_i^{(k)}}$. Conventionally, here we treat $\qubit{C^{(k)}_i}$ as the column vector $[((1/\sqrt{M})\omega_M^{k\cdot C_i(j)})_{j\in I_n}]^{t}$ and  $\bra{C^{(k)}_i}$ as the row vector $((1/\sqrt{M})\omega_M^{-k\cdot C_i(j)})_{j\in I_n}$. We use the notation $M_{k,C}$ to denote the matrix $(\ket{C_0^{(k)}},\ldots,\ket{C_{M(n)-1}^{(k)}})$, which equals 
\[
M_{k,C}=  \sum_{j\in I_n}\left(\frac{1}{\sqrt{M(n)}}\, 
\omega_{q}^{k\cdot C_0(M(n)-j\;\mathrm{mod}\,M(n))}\right)Q^j 
\]
and the conjugate transpose of $M_{k,C}$ can be expressed as 
\[
M_{k,C}^{\dag}=  \sum_{j\in I_n}\left(\frac{1}{\sqrt{M(n)}}\,
\omega_{q}^{k\cdot C_0(j)}\right)Q^j. 
\]
Clearly, these matrices are circulant since so are the matrices $(C_i(j))_{i,j\in I_n}$ and $(C_j(i))_{i,j\in I_n}$.
Therefore, as noted before, $M_{k,C}$ can be diagonalized by the quantum Fourier transform $F_M$ as follows:
\[
F_M^{-1} M_{k,C} F_M = \mathrm{diag} \left( \frac{1}{\sqrt{M}}\sum_{j\in I_n}\omega_{M}^{-ij}\omega_q^{k\cdot C_0(j)} \right)_{i\in I_n}.
\]
Similarly, we obtain the following diagonalization: 
\[
F_M^{-1} M_{k,C}^{\dag} F_M =  \mathrm{diag} \left( \frac{1}{\sqrt{M}}\sum_{j\in I_n}\omega_{M}^{ij}\omega_q^{- k\cdot C_0(j)} \right)_{i\in I_n}.
\]

\subsection{Proof of Theorem \ref{new-hardcore}(3)}

We shall give the proof of Theorem \ref{new-hardcore}(3). Our proof relies on the next lemma, in which we prove that, if we can approximate efficiently the matrix $F_MM_{k,C}F_M^{-1}$ (described in the previous subsection), we can construct efficiently a codeword-state decoder for $C$. The lemma requires the notion of {\em operator norm} $\|A\|$ of a complex square matrix $A$, defined as $\|A\|=\sup_{\qubit{\phi},\qubit{\psi}:\|\qubit{\phi}\|=\|\qubit{\psi}\|=1}|\bra{\phi}A\ket{\psi}|$.
 
\begin{lemma}\label{circulant-approx}
Let $C$ be an $(M(n),n)_{q(n)}$ circulant code. Let $k\in \field_{q(n)}^{+}$, $\delta\in[0,1]$, and let $D_{k}$ denote $F_{M(n)} M_{k,C}^{\dagger} F_{M(n)}^{-1}$. For each constant $k\in\field_{q(n)}^{+}$, let $\tilde{D}_k$ denote a linear operator such that $||\tilde{D}_{k}- D_{k}|| \leq \delta$, where $||\cdot ||$ denotes the operator norm.
If $\tilde{D}_{k}$ is computable in time polynomial in
$(n,q(n),\log_2M(n))$, then $C$ is $(1-\delta)^2$-quantum codeword-state decodable in time polynomial in $(n,q(n),\log_2M(n))$. 
\end{lemma}

\begin{proof}
Let $k\in\field_{q(n)}^{+}$. By omitting the script ``$n$,'' our desired codeword-state decoder $U_k$ that outputs $i$ from $\qubit{C_i^{(k)}}$ can be expressed in the form $F_{M}^{-1}\tilde{D}_{k} F_{M}$. Obviously, $U_k$ is a linear operator that can be realized in time polynomial 
in $(n,q,\log{M})$. 

Now, we wish to evaluate the success probability $|\bra{i}U_k\ket{C_i^{(k)}}|^2$ of obtaining $i$ by applying $U_k$ to $\ket{C_i^{(k)}}$. For convenience, let $\Delta_{k} = \tilde{D}_{k} - D_{k}$. This $\Delta_k$ satisfies the following inequality:
\[
|\bra{i}F_{M}^{-1}\Delta_{k}F_{M}\ket{C_i^{(k)}}|  
= |(\bra{i}F_{M}^{-1})\Delta_{k}(F_{M}\ket{C_i^{(k)}})|  
\leq \|\Delta_k\| = \|\tilde{D}_k - D_{k}\| \leq \delta.
\]
We then have 
\begin{eqnarray*}
|\bra{i}U_k\ket{C_i^{(k)}}| 
&=& \left|\bra{i}F_{M}^{-1}\tilde{D}_{k} F_{M}\ket{C_i^{(k)}}|\right| 
\;\;=\;\; \left|\bra{i}F_{M}^{-1}(D_{k}+\Delta_{k})F_{M}\ket{C_i^{(k)}}\right| \\
&\geq& \left|\bra{i}F_{M}^{-1}D_{k} F_{M}\ket{C_i^{(k)}}\right| - \left|\bra{i}F_{M}^{-1}\Delta_{k} F_{M}\ket{C_i^{(k)}}\right|,
\end{eqnarray*}
which is further bounded by
\begin{eqnarray*}
|\bra{i}U_k\ket{C_i^{(k)}}| 
&\geq& \left|\bra{i}M_{k,C}^{\dagger}\ket{C_i^{(k)}}\right| - ||\Delta_{k}|| 
\;\;\geq\;\; \left|\bra{i}M_{k,C}^\dag\ket{C_i^{(k)}}\right| - \delta\\
&=& \left|\braket{C_i^{(k)}}{C_i^{(k)}}\right| - \delta 
\;\;=\;\; 1-\delta.
\end{eqnarray*}
Thus, we can obtain $i$ from $\ket{C_i^{(k)}}$ with probability at least $(1-\delta)^2$. Since $\tilde{D}_k$ can be computed in time polynomial in $(n,q,\log{M})$, our codeword-state decode also runs in time polynomial in $(n,q,\log{M})$.
\end{proof}

With help of Lemma \ref{circulant-approx} together with Theorems \ref{code-hardcore} and \ref{negligible-code}, we prove Theorem \ref{new-hardcore}(3). 

\begin{proofof}{Theorem \ref{new-hardcore}(3)}
We wish to give a polynomial-time $(1-s)$-quantum codeword-state decoder for $\SLS^{p}$, where $s$ is a certain negligible function. Theorems \ref{code-hardcore} and \ref{negligible-code}
then guarantee the quantum hardcore property of $\SLS^{p}$. 

Let fix $n\in\nat$ and consider a new code $C$ defined as follows: let $C_i(j)$ be $\SLS^{p}_{-i}(j)$ (using ``$-i$'' instead of ``$i$'') for each pair $i,j\in I_n$. Since $C$ is a circulant code, we hereafter consider its associated matrix $M_{2,C}^{\dagger} = \sum_{j\in \field_{p}}\left( (1/\sqrt{p})\omega_{2}^{C_0(j)} \right)Q^j$.

To obtain a quantum codeword-state decoder for $C_i$, we use Lemma \ref{circulant-approx}. First, we define a useful constant $c_p$ as follows: $c_p = 1$ if $p\equiv 1 \bmod 4$, and $c_p = \iota$ (\ie the unit of imaginary numbers) if $p\equiv 3 \bmod 4$. This constant $c_p$ satisfies the 
 following equation (see e.g. \cite{CP01}):
\[
 (*) \hs{10} \frac{1}{\sqrt{p}}\sum_{j\in\field_{p}} \legendre{j}{p}\omega_p^{a\cdot j} = c_p\legendre{a}{p}
\]
for any number $a\in[0,p-1]_{\integer}$. Let $D_2= F_p^{-1}M_{2,C}^{\dagger}F_p$, which equals
\[
D_2 = \mathrm{diag} \left( \frac{1}{\sqrt{p}}\sum_{j\in \field_{p}}\omega_{p}^{ij}\omega_2^{- C_0(j)} \right)_{i\in I_n} = \mathrm{diag} \left( \frac{1}{\sqrt{p}} +  \frac{1}{\sqrt{p}}\sum_{j\in \field_{p}}\left(\frac{j}{p} \right) \omega_p^{ij} \right)_{i\in I_n}
\]
because $\omega_2^{-C_0(0)} = 1$ and $\omega_2^{-C_0(j)} = \left( \frac{j}{p}\right)$ for any number $j\in\field_{p}^{+}$. By (*), we have
\[
D_2 = \mathrm{diag} \left( \frac{1}{\sqrt{p}} + c_p\left(\frac{i}{p} \right) \right)_{i\in I_n} = \mathrm{diag}\left( \frac{1}{\sqrt{p}},\frac{1}{\sqrt{p}}+c_p\omega_2^{-C_0(1)},\ldots, \frac{1}{\sqrt{p}}+c_p\omega_2^{-C_0(p-1)} \right).
\]

We define our desired linear operator $\tilde{D}_2$ as 
$
\tilde{D}_2 = \mathrm{diag} \left(  0, c_p \omega_2^{-C_0(1)}, \ldots, c_p\omega_2^{-C_0(p-1)} \right).
$
This definition makes the operator norm $\|D_2-\tilde{D}_2\|$ equal 
\[
 \|D_2-\tilde{D}_2\| = \left\| \mathrm{diag} \left(  \frac{1}{\sqrt{p}}, \ldots, \frac{1}{\sqrt{p}} \right) \right\| 
= \frac{1}{\sqrt{p}}.
\]
How can we realize this $\tilde{D}_2$? The operator $\tilde{D}_2$ can be realized by the following polynomial-time algorithm. 
\begin{quote}\vs{-1}
On input $\qubit{i}\qubit{0}$, where $i\in\field_q$, compute $c_p\qubit{i}\qubit{C_0(i)}$ in a reversible fashion. Apply the phase-shift transform that changes $\qubit{i}\qubit{a}$ to $\omega_2^{-a}\qubit{i}\qubit{a}$. Uncompute $\qubit{C_0(i)}$ in the last register and we obtain $c_p\omega_2^{-C_0(i)}\qubit{i}\qubit{0}$. Finally, when $i=0$, we reject the input.
\end{quote}\vs{-1}
Therefore, Lemma \ref{circulant-approx} gives a $(1-1/\sqrt{p})^2$-quantum codeword-state decoder for $C$ that runs in time polynomial in $n$. Since $(1-1/\sqrt{p})^2\geq 1-2/\sqrt{p}$, it suffices to define $s(n)=2/\sqrt{p}$. 

To list-decode $\SLS$, since $\SLS^{p}_{i}(j) = C_{-i}(j)$, we first find $-i$ from the codeword $C_{-i}(\cdot)$ and then output $i$. This procedure gives rise to a quantum list-decoder for $\SLS^{p}$.
\end{proofof}

\bs
\paragraph{Acknowledgments:}
The first author is grateful to Harumichi Nishimura for his pointing out an early error.


\bs
\section*{Appendix A: Proof of Lemma \ref{fidelity}}

We give the proof of Lemma \ref{fidelity}.
Fix $n$ arbitrarily and drop the subscript ``$n$'' for simplicity. For each index $\ell\in[0,q-1]_{\integer}$, we define  $d^{(k)}_{\ell}(C_x,C_y)=|\{r\in I_n \mid k(C_x(r)-C_y(r))=\ell\;\mathrm{mod}\;q\}|$. Since  $F(\qubit{C^{(k)}_x},\qubit{C^{(k)}_y})=|\braket{C^{(k)}_x}{C^{(k)}_y}|$, it follows that
\begin{eqnarray*}
F(\qubit{C^{(k)}_x},\qubit{C^{(k)}_y})
&=& \left| \frac{1}{M}\cdot\omega_q^{0}\cdot d_{0}^{(k)}(C_x,C_y)  
+ \frac{1}{M} \sum_{\ell=1}^{q-1}\omega_q^{\ell}\cdot d^{(k)}_{\ell}(C_x,C_y) \right| \\
&\geq& \left| 1 - \frac{d(C_x,C_y)}{M} \right| 
- \frac{1}{M} \sum_{\ell=1}^{q-1}\left| \omega_q^{\ell}\cdot d^{(k)}_{\ell}(C_x,C_y) \right| \\
&=& (1 - \Delta(C_x,C_y) ) - \Delta(C_x,C_y) 
\;\;=\;\; 1 - 2\Delta(C_x,C_y), 
\end{eqnarray*}
which gives the desired bound of the lemma.
In particular, when $q=2$, since $d^{(k)}_1(C_x,C_y)=d(C_x,C_y)$ and $\omega_2=-1$, we obtain the equality $F(\qubit{C^{(k)}_x},\qubit{C^{(k)}_y}) =  1 - 2\Delta(C_x,C_y)$.

\section*{Appendix B: Proof of Lemma \ref{johnson-bound}}

For the proof of Lemma \ref{johnson-bound}, we need to elaborate the brief description given in Section \ref{sec:hardcore-property} on an interpretation of {\em presence}. 

For readability, we fix $n$ and omit this $n$ (for example, we write ``$q$'' instead of ``$q(n)$'') in the following proof. Let $M=|I_n|$. 
Let $v$ be any quantumly corrupted word that $\tilde{O}$ represents. We view this $v$ as the real vector $(v[0],v[1],\ldots,v[M-1])$ in the
$Mq$ dimensional real space defined as follows: let $v[r]$, the $r$th block of $v$, be $( |\alpha_{r,0}|^2, |\alpha_{r,1}|^2, \ldots, |\alpha_{r,q-1}|^2)$ 
if $\tilde{O}\qubit{r}\qubit{0}\qubit{0^{\ell(n)}} = \sum_{z\in \integer_{q}} \alpha_{r,z}\qubit{r}\qubit{z}\qubit{\phi_{r,z}}$. Let $\{C_1,C_2, \ldots, C_m\}$ be the set of all codewords that lie
``close'' to the given quantumly corrupted word. For each $C_i$, we define $c_i$ to be the corresponding vector
defined as follows: let $c_i[r]$, the $r$th block of $c_i$, consist of zeros and one $1$ at the $z$th component if $C_i(r) = z$. 

Let $i\in [m]$ be any index. We further introduce a new parameter $\beta\in [0, 1]$ and define
$w = \beta\cdot v + \frac{1-\beta}{q}\cdot \vec{1}$, where $\vec{1}$ is the vector of all $1$s. Note that $\braket{\vec{1}}{ \vec{1}} = Mq$. Note also that the (Hamming)
distance $d(C_i,C_j)$ between codewords $C_i$ and $C_j$ is lower-bounded by $d$. Moreover, we have $\braket{c_i}{v} = \sum_{r} |\alpha_{r,C_i(r)}|^2 =
M\cdot \Pre_{\tilde{O}}(C_i)$, where $\braket{v}{w}$ denotes the standard inner product of two vectors $v$ and $w$. Note that $\braket{c_i}{c_i}=M$. 

For each $i\in[m]$, let $\hat{c}_i = c_i - w$. We consider the set $\{\hat{c}_i\}_{i\in[m]}$. We define the space $\KK = 
\{x\in\real^{M}\times\real^{q} \mid \forall r\in I_n[\sum_{z\in\integer_{q}} x_{r,z} = 0]\}$. Notice that $\{\hat{c}_i\}_{i\in[m]}\subseteq \KK$ and $dim(\KK) = M(q -1)$.
Moreover, let $\hat{w}$ be the projection of $w$ onto $\KK$. 

Let us first describe the following result stated in \cite{GS00b}.

\begin{appendixlemma}\label{Sudan-lemma}{\rm \cite{GS00b}}
Let $\{u_i\}_{i\in [m]}\subseteq \real^{K}$ be $m$ non-zero vectors such that $\braket{u_i}{u_j}\leq 0$ for any distinct pair $i,j\in[m]$. Let $\gamma>0$.
\begin{enumerate}
\item If $\exists v\in\real^{K}\forall i\in[m][\braket{v}{u_i}>0]$, 
then $m\leq K$.
\vs{-1}
\item If $\forall i\in[m][\|u_i\|=1]$ and $\forall i,j\in[m][i\neq j\rightarrow \braket{u_i}{u_j}\leq -\gamma]$, then $m\leq 1+1/\gamma$.
\vs{-2}
\item If $\exists v\in\real^{K}\forall i\in[m][\braket{v}{u_i}\geq0]$, 
then $m\leq 2K-1$.
\end{enumerate}
\end{appendixlemma}

\paragraph{The First Upper Bound.}
Let $(i,j)$ be any distinct pair taken from $[m]$. The values $\braket{c_i}{w}$, $\braket{w}{w}$, and $\braket{c_i}{c_j}$ can be bounded as follows:
\begin{eqnarray*}
&& \braket{c_i}{w} = \beta\braket{c_i}{v} +
\frac{1 - \beta}{q} \braket{c_i}{\vec{1}} = \beta M \Pre_{\tilde{O}}(C_i) +
\frac{1 - \beta}{q} M \geq  \beta M \left( \frac{1}{q} + \varepsilon\right) 
+ \frac{M(1 - \beta)}{q}. \\
&&
\braket{w}{w} = \beta^2 \braket{v}{v} +
\frac{2\beta(1 - \beta)}{q} \braket{v}{\vec{1}} +
\frac{(1 - \beta)^2}{q^2} \braket{\vec{1}}{\vec{1}} \leq M\beta^2 +
\frac{2M\beta(1 - \beta)}{q} + \frac{M(1 - \beta)^2}{q}. \\
&&
\braket{c_i}{c_j} = M - d(C_i,C_j) \leq M - d.
\end{eqnarray*}
Let us consider the set $\{\hat{c}_i\mid i\in [m]\}$. Now, the inner product $\braket{\hat{c}_i}{\hat{c}_j}$ for a distinct pair $i,j\in[m]$ is estimated as
\begin{eqnarray*}
\braket{\hat{c}_i}{\hat{c}_j} &=& 
\braket{c_i}{c_j} +  \braket{w}{w} - \braket{c_i}{w}  - \braket{c_j}{w} \\ 
&\leq&  M - d + M\beta^2 + 
\frac{2N\beta(1 - \beta)}{q} + \frac{M(1 - \beta)^2}{q} - 2M \left[ \left( \frac{1}{q} + \varepsilon\right) \beta +
\frac{1 - \beta}{q} \right] \\ 
&=& M \left( 1 - \frac{1}{q} \right) \left[ \beta^2 -
\frac{2q\varepsilon}{q - 1} \beta + 1\right] - d.
\end{eqnarray*}
For our convenience, we write $d = \left(1 - \frac{1}{q}\right) (1 - \delta)M$ using  an appropriate value  $\delta\in[0, 1]$. It thus follows that:
\begin{eqnarray*}
\braket{\hat{c}_i}{\hat{c}_j} &\leq&  
M \left( 1 - \frac{1}{q} \right) \left[ \beta^2 -
\frac{2q\varepsilon}{q - 1} \beta + 1\right] - \left( 1 -
\frac{1}{q} \right)(1 - \delta) M \\
&=& M \left( 1 - \frac{1}{q} \right) \left[ \beta^2 -
\frac{2q\varepsilon}{q - 1} \beta + \delta\right].
\end{eqnarray*}
To apply Lemma \ref{Sudan-lemma}(1), we want to make $\braket{\hat{c}_i}{\hat{c}_j} < 0$. To do so, we require that $\beta^2 - \frac{2q\varepsilon}{q-1} \beta + \delta < 0$, which is equivalent
to  $\varepsilon  > \frac{1}{2} \left( 1 - \frac{1}{q} \right) \left( \beta + \frac{\delta}{\beta}\right)$. To minimize the value $\beta + \frac{\delta}{\beta}$, it suffices to take $\beta = \sqrt{\delta}$. By replacing
$\beta$ by $\sqrt{\delta}$, we obtain $\varepsilon  > \sqrt{\delta}\left( 1 - \frac{1}{q}\right)$. Since $\delta = 1 - \frac{d}{M}\left( 1 + \frac{1}{q-1}\right)$, we obtain the bound
$\varepsilon  > \left(1 - \frac{1}{q}\right) \sqrt{1 - \frac{d}{M} \left( 1 + \frac{1}{q-1}\right)}$.

Since $\hat{c}_i,\hat{w} \in\KK$, we have $\braket{\hat{c}_i}{\hat{w}} = \braket{\hat{c}_i}{w}$. It thus follows that 
\[
\braket{\hat{c}_i}{\hat{w}}  = \braket{c_i-w}{w} = \braket{c_i}{w} - \braket{w}{w} 
\geq
\frac{M\beta}{q} \left[ q \varepsilon - \beta(q - 1)\right].
\]
Since $\beta = \sqrt{\delta}$, we have
\[
\braket{\hat{c}_i}{\hat{w}} >
\frac{N\sqrt{\delta}}{q} \left[ q\sqrt{\delta}\left( 1 -
\frac{1}{q} \right) -
\sqrt{\delta}(q - 1) \right] = 0.
\]
This implies, by Lemma \ref{Sudan-lemma}(1), that $m \leq dim(\KK) = M(q - 1)$.

\paragraph{The Second Upper Bound.}

\sloppy We show the second upper bound. 
Recall that $\braket{\hat{c}_i}{\hat{c}_j} \leq M \left( 1 - \frac{1}{q} \right) \left[ \beta^2 - \frac{2q\varepsilon}{q-1} \beta + \delta \right]$. We
choose $\beta = \frac{q \varepsilon}{q-1}$ so that $\beta^2 - \frac{2q\varepsilon}{q-1}\beta  = -\beta^2$. If $\delta < \beta^2 =  \left(\frac{q \varepsilon}{q-1} \right)^2$, then clearly, we have $\braket{\hat{c}_i}{\hat{c}_j} \leq
M \left( 1 - \frac{1}{q} \right) \left(\delta - \beta^2 \right) < 0$. Note that the condition $\delta <  \left(\frac{q\varepsilon}{q-1}\right)^2$
is equivalent to  $\varepsilon > \sqrt{\delta}\left(1 - \frac{1}{q}\right)$. Since
$\delta = 1 - \frac{d}{M}\left(1+\frac{1}{q-1} \right) = 1- \frac{q d}{M(q-1)}$, we obtain that $\varepsilon  > \left(1 - \frac{1}{q} \right) \sqrt{ 1 - \frac{d}{M} \left(1 + \frac{1}{q-1}\right)}$ as before. 
Since $\sum_{i=1}^{n}a_i^2\geq (1/n)\left(\sum_{i=1}^{n}a_i\right)^2$ for any real-valued series $\{a_i\}_{i\in[n]}$, we obtain that $\braket{v}{v} = \sum_{r}\sum_{i}|\alpha_{r,i}|^4\geq M/q$. Thus, $\braket{w}{w}$ is lower-bounded by 
\[
\braket{w}{w} \geq \frac{M\beta^2}{q} + \frac{2M(1-\beta)}{q} + \frac{M(1-\beta)^2}{q}
= \frac{M}{q}
\]
Since $\Pre_{\tilde{O}}(C_i)\leq 1$, we obtain $\braket{c_i}{w}\leq \beta M + M(1-\beta)/q$. 
We also have
\[
\|\hat{c}_i\|^2 
= \braket{c_i}{c_i} + \braket{w}{w} - 2\braket{c_i}{w} 
\geq M + \frac{M}{q} - 2\left[\beta M + \frac{M(1-\beta)}{q}\right]
= M\left(1-\frac{1}{q}\right)(1-\beta)
\]
By normalizing $\hat{c}_i$, we write $u_i = \frac{\hat{c}_i}{\| \hat{c}_i \|}$. Hence, we have 
\[
\braket{u_i}{u_j} = 
\frac{\braket{ \hat{c}_i}{\hat{c}_j}}{\|\hat{c}_i \|\cdot \|\hat{c}_j \|}
\leq \frac{M(1-1/q)(\delta-\beta^2)}{M(1-1/q)(1-\beta)} = 
 - \frac{\beta^2 - \delta}{1 - \beta}.
\]
By Lemma \ref{Sudan-lemma}(2), since $1-1/q-\varepsilon\geq0$, we conclude that
\[
m \leq 1 + \frac{1 - \beta}{\beta^2 - \delta} =
\frac{\beta^2-\beta-\delta+1}{\beta^2 - \delta} 
\leq 
\frac{d \left(1 - \frac{1}{q} \right)}{M\varepsilon^2 + \left(1 - \frac{1}{q} \right)\left[d- M \left(1 - \frac{1}{q} \right) \right]}.
\]

\paragraph{The Equality Case.}

Assume that $\varepsilon = \left( 1 - \frac{1}{q} \right)\sqrt{1 - \frac{d}{M} \left( 1 + \frac{1}{q-1}\right)}$, which is equivalent to   $\varepsilon = \sqrt{\delta} \left(1 - \frac{1}{q}\right)$. We
want to show that $m \leq  2M(q-1)-1$ by applying Lemma \ref{Sudan-lemma}(3). Recall that $\braket{\hat{c}_i}{\hat{c}_j } \leq M \left(1 - \frac{1}{q} \right) \left[ \beta^2 - \frac{2q \varepsilon}{q-1} \beta + \delta\right]$.
Taking $\beta = \frac{q \varepsilon}{q-1}$ ($= \sqrt{\delta}$), we immediately obtain $\braket{\hat{c}_i }{\hat{c}_j } \leq 0$. Let us consider $\hat{w}\in\KK$. Since $\braket{\hat{c}_i}{\hat{w}} = \braket{\hat{c}_i}{w}$, it follows that 
\begin{eqnarray*}
\braket{\hat{c}_i}{\hat{w}} &=& \braket{c_i}{w} - \braket{w}{w} \\ 
&\geq& \beta M \left( \frac{1}{q} + \varepsilon \right) + \frac{M(1 - \beta)}{q} - N\beta^2 - \frac{2M\beta(1 - \beta)}{q} - \frac{M(1 - \beta)^2}{q} \\
&=& \frac{M}{q} \left[\beta(1 + q \varepsilon) + (1 - \beta) - q\beta^2 - \beta(1 - \beta) - (1 - \beta)^2 \right].
\end{eqnarray*}
Replacing $\beta$ and $\varepsilon$  by the appropriate terms using $\delta$, we have
\[
\braket{\hat{c}_i }{\hat{w}} \geq \frac{M}{q} \left[ \left( 1 + q\cdot \frac{\sqrt{\delta} (q - 1)}{q} \right) \sqrt{\delta} - q\delta \right] = \frac{M}{q} \left[\sqrt{\delta} - \delta \right] \geq 0
\]
because $0 \leq \delta \leq 1$.
Hence, by Lemma \ref{Sudan-lemma}(3), we obtain that $m = 2M(q - 1) - 1$.

\end{document}